%% file: main.tex
\documentclass{article}
\usepackage[preprint]{neurips_2025}
\makeatletter\renewcommand{\@notice}{}\makeatother
\AtBeginDocument{\newgeometry{textwidth=6.5in,textheight=9in,top=1in,headheight=12pt,headsep=25pt,footskip=30pt}}
\usepackage[T1]{fontenc}
\usepackage[utf8]{inputenc}
\usepackage{microtype}
\usepackage{amsmath}
\usepackage{graphicx}
\usepackage{booktabs}\usepackage{array}
\usepackage{adjustbox}
\usepackage{float}\usepackage{wrapfig}
\usepackage[section]{placeins}
\usepackage{xcolor}
\usepackage{enumitem}
\setcitestyle{square,authoryear,citesep={,}}
\usepackage{hyperref}\hypersetup{pdftitle={AircraftWitness: How Often Aircraft Can Locate Recordings},pdfauthor={Kris Atallah}}
\usepackage{cleveref}\usepackage{xurl}
\crefformat{section}{\S#2#1#3}\Crefformat{section}{\S#2#1#3}\crefformat{subsection}{\S#2#1#3}\Crefformat{subsection}{\S#2#1#3}\crefrangeformat{section}{\S\S#3#1#4 to #5#2#6}
\crefname{figure}{Fig.}{Figs.}\Crefname{figure}{Fig.}{Figs.}\crefname{table}{Table}{Tables}\Crefname{table}{Table}{Tables}\crefname{equation}{Eq.}{Eqs.}
\crefname{appendix}{Appendix}{Appendices}\Crefname{appendix}{Appendix}{Appendices}
\makeatletter
\newif\ifmissing \newif\iftrial
\DeclareRobustCommand{\missing}[1]{\iftrial\global\missingtrue\else\textcolor{red}{[#1]}\PackageWarning{numbers}{missing #1 outside opt}\fi}
\newcommand{\tbd}[1]{\missing{tbd}}
\DeclareRobustCommand{\opt}[1]{\global\missingfalse\setbox0=\hbox{\trialtrue#1}\ifmissing\expandafter\ignorespaces\else#1\fi}
\newcommand{\figw}{\linewidth}\newcommand{\optfig}[4][t]{\ifcsname figok@#2\endcsname\begin{figure}[#1]\centering\includegraphics[width=\figw]{figs/#2.pdf}\caption{#3}\label{#4}\end{figure}\fi}
\newcommand{\ifhave}[3]{\ifcsname ok@#1\endcsname#2\else#3\fi}
\newcommand{\iffig}[3]{\ifcsname figok@#1\endcsname#2\else#3\fi}
\makeatother
\input{build/numbers.tex}
\input{build/figready.tex}

\title{AircraftWitness: How Often Aircraft\\ Can Locate Recordings}
\IfFileExists{build/authors.tex}{\input{build/authors.tex}}{\input{authors.tex}}
\date{}

\begin{document}
\maketitle
\input{sec/abstract}
\input{sec/intro}
\input{sec/related}
\input{sec/method}
\input{sec/results}
\input{sec/meaning}
{\renewcommand{\FloatBarrier}{}\input{sec/ethics}}%
\input{sec/conclusion}
\input{sec/statements}
\bibliographystyle{aw}\bibliography{refs}
\newpage
\appendix\raggedbottom\renewcommand{\FloatBarrier}{}\renewcommand{\figw}{\linewidth}%
\input{sec/appendix}
\end{document}

%% file: build/numbers.tex
\newcommand{\relUrl}{\url{https://github.com/nyu-kr/aircraftwitness}}%
\expandafter\def\csname ok@relUrl\endcsname{}%
\newcommand{\dateFJet}{250}%
\expandafter\def\csname ok@dateFJet\endcsname{}%
\newcommand{\baseLivN}{37}%
\expandafter\def\csname ok@baseLivN\endcsname{}%
\newcommand{\baseLivRms}{4.7}%
\expandafter\def\csname ok@baseLivRms\endcsname{}%
\newcommand{\baseLivOut}{3}%
\expandafter\def\csname ok@baseLivOut\endcsname{}%
\newcommand{\baseLivOutKm}{6.6}%
\expandafter\def\csname ok@baseLivOutKm\endcsname{}%
\newcommand{\clockClips}{38}%
\expandafter\def\csname ok@clockClips\endcsname{}%
\newcommand{\clockPassSd}{1.1}%
\expandafter\def\csname ok@clockPassSd\endcsname{}%
\newcommand{\clockOffsetErr}{0.69}%
\expandafter\def\csname ok@clockOffsetErr\endcsname{}%
\newcommand{\clockDriftErr}{0.25}%
\expandafter\def\csname ok@clockDriftErr\endcsname{}%
\newcommand{\clockLivNatOff}{-2.0}%
\expandafter\def\csname ok@clockLivNatOff\endcsname{}%
\newcommand{\clockLivNatSd}{0.9}%
\expandafter\def\csname ok@clockLivNatSd\endcsname{}%
\newcommand{\unhBThreeK}{2}%
\expandafter\def\csname ok@unhBThreeK\endcsname{}%
\newcommand{\unhBThreeN}{360}%
\expandafter\def\csname ok@unhBThreeN\endcsname{}%
\newcommand{\clockBudAoOff}{1.7}%
\expandafter\def\csname ok@clockBudAoOff\endcsname{}%
\newcommand{\clockLag}{0.27}%
\expandafter\def\csname ok@clockLag\endcsname{}%
\newcommand{\clockEsDays}{5.1}%
\expandafter\def\csname ok@clockEsDays\endcsname{}%
\newcommand{\clockEsGated}{588}%
\expandafter\def\csname ok@clockEsGated\endcsname{}%
\newcommand{\clockEsInliers}{86}%
\expandafter\def\csname ok@clockEsInliers\endcsname{}%
\newcommand{\clockEsDrift}{1.31}%
\expandafter\def\csname ok@clockEsDrift\endcsname{}%
\newcommand{\clockEsDriftSd}{0.09}%
\expandafter\def\csname ok@clockEsDriftSd\endcsname{}%
\newcommand{\clockEsAnchor}{$-$0.38}%
\expandafter\def\csname ok@clockEsAnchor\endcsname{}%
\newcommand{\clockEsAnchorSd}{0.23}%
\expandafter\def\csname ok@clockEsAnchorSd\endcsname{}%
\newcommand{\clockGapN}{10}%
\expandafter\def\csname ok@clockGapN\endcsname{}%
\newcommand{\clockGapR}{0.89}%
\expandafter\def\csname ok@clockGapR\endcsname{}%
\newcommand{\defDpLivN}{19}%
\expandafter\def\csname ok@defDpLivN\endcsname{}%
\newcommand{\defDpLivNoneK}{17}%
\expandafter\def\csname ok@defDpLivNoneK\endcsname{}%
\newcommand{\defDpLivOffK}{1}%
\expandafter\def\csname ok@defDpLivOffK\endcsname{}%
\newcommand{\defDpLivMixK}{8}%
\expandafter\def\csname ok@defDpLivMixK\endcsname{}%
\newcommand{\defDpLivRnK}{7}%
\expandafter\def\csname ok@defDpLivRnK\endcsname{}%
\newcommand{\defDpLivDfnK}{10}%
\expandafter\def\csname ok@defDpLivDfnK\endcsname{}%
\newcommand{\flAdsbxDays}{10}%
\expandafter\def\csname ok@flAdsbxDays\endcsname{}%
\newcommand{\flAdsbxRatio}{2.2}%
\expandafter\def\csname ok@flAdsbxRatio\endcsname{}%
\newcommand{\glInusN}{110}%
\expandafter\def\csname ok@glInusN\endcsname{}%
\newcommand{\glInusKm}{7}%
\expandafter\def\csname ok@glInusKm\endcsname{}%
\newcommand{\glInusExpAll}{0.9}%
\expandafter\def\csname ok@glInusExpAll\endcsname{}%
\newcommand{\glIneuN}{88}%
\expandafter\def\csname ok@glIneuN\endcsname{}%
\newcommand{\glIneuK}{none}%
\expandafter\def\csname ok@glIneuK\endcsname{}%
\newcommand{\glAuInusN}{110}%
\expandafter\def\csname ok@glAuInusN\endcsname{}%
\newcommand{\glAuInusK}{3}%
\expandafter\def\csname ok@glAuInusK\endcsname{}%
\newcommand{\glAuInusDec}{none}%
\expandafter\def\csname ok@glAuInusDec\endcsname{}%
\newcommand{\glAuIneuN}{88}%
\expandafter\def\csname ok@glAuIneuN\endcsname{}%
\newcommand{\glAuIneuK}{2}%
\expandafter\def\csname ok@glAuIneuK\endcsname{}%
\newcommand{\glAuIneuDec}{2}%
\expandafter\def\csname ok@glAuIneuDec\endcsname{}%
\newcommand{\glAuXceuN}{304}%
\expandafter\def\csname ok@glAuXceuN\endcsname{}%
\newcommand{\glAuXceuK}{4}%
\expandafter\def\csname ok@glAuXceuK\endcsname{}%
\newcommand{\glAuXceuDec}{3}%
\expandafter\def\csname ok@glAuXceuDec\endcsname{}%
\newcommand{\glAuXceuDecConf}{2}%
\expandafter\def\csname ok@glAuXceuDecConf\endcsname{}%
\newcommand{\glAuXceuTwoN}{255}%
\expandafter\def\csname ok@glAuXceuTwoN\endcsname{}%
\newcommand{\aoSites}{5}%
\expandafter\def\csname ok@aoSites\endcsname{}%
\newcommand{\aoResolvedN}{none}%
\expandafter\def\csname ok@aoResolvedN\endcsname{}%
\newcommand{\clockLongSites}{28}%
\expandafter\def\csname ok@clockLongSites\endcsname{}%
\newcommand{\clockLongFit}{9}%
\expandafter\def\csname ok@clockLongFit\endcsname{}%
\newcommand{\clockLongRes}{none}%
\expandafter\def\csname ok@clockLongRes\endcsname{}%
\newcommand{\clockLongLo}{$-$12}%
\expandafter\def\csname ok@clockLongLo\endcsname{}%
\newcommand{\clockLongHi}{+23}%
\expandafter\def\csname ok@clockLongHi\endcsname{}%
\newcommand{\clockLongChi}{1{,}200}%
\expandafter\def\csname ok@clockLongChi\endcsname{}%
\newcommand{\clockLongDof}{8}%
\expandafter\def\csname ok@clockLongDof\endcsname{}%
\newcommand{\clockLongHeadDec}{7}%
\expandafter\def\csname ok@clockLongHeadDec\endcsname{}%
\newcommand{\clockLongHeadAgree}{3}%
\expandafter\def\csname ok@clockLongHeadAgree\endcsname{}%
\newcommand{\clockLongBlocks}{65}%
\expandafter\def\csname ok@clockLongBlocks\endcsname{}%
\newcommand{\clockLongBlocksRes}{12}%
\expandafter\def\csname ok@clockLongBlocksRes\endcsname{}%
\newcommand{\clockLongCtl}{3.00}%
\expandafter\def\csname ok@clockLongCtl\endcsname{}%
\newcommand{\clockLongCtlN}{3}%
\expandafter\def\csname ok@clockLongCtlN\endcsname{}%
\newcommand{\eclN}{187}%
\expandafter\def\csname ok@eclN\endcsname{}%
\newcommand{\eclK}{5}%
\expandafter\def\csname ok@eclK\endcsname{}%
\newcommand{\eclSites}{38}%
\expandafter\def\csname ok@eclSites\endcsname{}%
\newcommand{\eclSitesHit}{3}%
\expandafter\def\csname ok@eclSitesHit\endcsname{}%
\newcommand{\eclClockInTask}{9}%
\expandafter\def\csname ok@eclClockInTask\endcsname{}%
\newcommand{\eclClockHit}{1}%
\expandafter\def\csname ok@eclClockHit\endcsname{}%
\newcommand{\eclCmpDgN}{86}%
\expandafter\def\csname ok@eclCmpDgN\endcsname{}%
\newcommand{\eclCmpDgK}{1}%
\expandafter\def\csname ok@eclCmpDgK\endcsname{}%
\newcommand{\eclCmpDgOursK}{3}%
\expandafter\def\csname ok@eclCmpDgOursK\endcsname{}%
\newcommand{\eclCmpDpSrcK}{4}%
\expandafter\def\csname ok@eclCmpDpSrcK\endcsname{}%
\newcommand{\eclCmpDpPypiK}{3}%
\expandafter\def\csname ok@eclCmpDpPypiK\endcsname{}%
\newcommand{\eclTTwoSiteDays}{84}%
\expandafter\def\csname ok@eclTTwoSiteDays\endcsname{}%
\newcommand{\eclTTwoK}{none}%
\expandafter\def\csname ok@eclTTwoK\endcsname{}%
\newcommand{\benchYrN}{155}%
\expandafter\def\csname ok@benchYrN\endcsname{}%
\newcommand{\benchYrK}{2}%
\expandafter\def\csname ok@benchYrK\endcsname{}%
\newcommand{\benchYrP}{0.92}%
\expandafter\def\csname ok@benchYrP\endcsname{}%
\newcommand{\benchYrBridgeSame}{22}%
\expandafter\def\csname ok@benchYrBridgeSame\endcsname{}%
\newcommand{\benchYrBridgeN}{34}%
\expandafter\def\csname ok@benchYrBridgeN\endcsname{}%
\newcommand{\benchYrBridgePred}{30}%
\expandafter\def\csname ok@benchYrBridgePred\endcsname{}%
\newcommand{\aoLocN}{295}%
\expandafter\def\csname ok@aoLocN\endcsname{}%
\newcommand{\aoLocSites}{5}%
\expandafter\def\csname ok@aoLocSites\endcsname{}%
\newcommand{\aoLocTopSiteK}{17}%
\expandafter\def\csname ok@aoLocTopSiteK\endcsname{}%
\newcommand{\aoLocBTwoN}{61}%
\expandafter\def\csname ok@aoLocBTwoN\endcsname{}%
\newcommand{\aoLocBTwoSites}{2}%
\expandafter\def\csname ok@aoLocBTwoSites\endcsname{}%
\newcommand{\aoLocBTwoAns}{6}%
\expandafter\def\csname ok@aoLocBTwoAns\endcsname{}%
\newcommand{\aoSlackK}{24}%
\expandafter\def\csname ok@aoSlackK\endcsname{}%
\newcommand{\aoSlackExp}{2.3}%
\expandafter\def\csname ok@aoSlackExp\endcsname{}%
\newcommand{\aoShiftK}{none}%
\expandafter\def\csname ok@aoShiftK\endcsname{}%
\newcommand{\aoShiftMin}{5}%
\expandafter\def\csname ok@aoShiftMin\endcsname{}%
\newcommand{\aoFullShiftK}{1}%
\expandafter\def\csname ok@aoFullShiftK\endcsname{}%
\newcommand{\aoBThreeN}{360}%
\expandafter\def\csname ok@aoBThreeN\endcsname{}%
\newcommand{\aoBThreeRiskPct}{6}%
\expandafter\def\csname ok@aoBThreeRiskPct\endcsname{}%
\newcommand{\aoBThreeK}{21}%
\expandafter\def\csname ok@aoBThreeK\endcsname{}%
\newcommand{\aoBThreeAns}{98}%
\expandafter\def\csname ok@aoBThreeAns\endcsname{}%
\newcommand{\aoBThreeNoMatch}{241}%
\expandafter\def\csname ok@aoBThreeNoMatch\endcsname{}%
\newcommand{\aoBThreeNoPass}{21}%
\expandafter\def\csname ok@aoBThreeNoPass\endcsname{}%
\newcommand{\aoBThreeMedKm}{4.6}%
\expandafter\def\csname ok@aoBThreeMedKm\endcsname{}%
\newcommand{\aoBThreeSites}{23}%
\expandafter\def\csname ok@aoBThreeSites\endcsname{}%
\newcommand{\aoBThreeSitesHit}{13}%
\expandafter\def\csname ok@aoBThreeSitesHit\endcsname{}%
\newcommand{\aoBThreeShiftExp}{0.8}%
\expandafter\def\csname ok@aoBThreeShiftExp\endcsname{}%
\newcommand{\aoBThreeShiftP}{$2\times10^{-15}$}%
\expandafter\def\csname ok@aoBThreeShiftP\endcsname{}%
\newcommand{\aoBThreeOutSitesHit}{12}%
\expandafter\def\csname ok@aoBThreeOutSitesHit\endcsname{}%
\newcommand{\aoBThreeOutK}{18}%
\expandafter\def\csname ok@aoBThreeOutK\endcsname{}%
\newcommand{\aoBThreeHitEdge}{none}%
\expandafter\def\csname ok@aoBThreeHitEdge\endcsname{}%
\newcommand{\aoBThreeOnePassK}{18}%
\expandafter\def\csname ok@aoBThreeOnePassK\endcsname{}%
\newcommand{\aoBThreeCtlMaxK}{2}%
\expandafter\def\csname ok@aoBThreeCtlMaxK\endcsname{}%
\newcommand{\aoBThreeDpK}{29}%
\expandafter\def\csname ok@aoBThreeDpK\endcsname{}%
\newcommand{\aoBThreeDpPct}{8}%
\expandafter\def\csname ok@aoBThreeDpPct\endcsname{}%
\newcommand{\aoBThreeAnsPerHit}{5}%
\expandafter\def\csname ok@aoBThreeAnsPerHit\endcsname{}%
\newcommand{\aoBThreeDpBoth}{14}%
\expandafter\def\csname ok@aoBThreeDpBoth\endcsname{}%
\newcommand{\aoBThreePsN}{23}%
\expandafter\def\csname ok@aoBThreePsN\endcsname{}%
\newcommand{\aoBThreePsK}{1}%
\expandafter\def\csname ok@aoBThreePsK\endcsname{}%
\newcommand{\aoBThreePsExp}{0.18}%
\expandafter\def\csname ok@aoBThreePsExp\endcsname{}%
\newcommand{\aoBThreePsZeroPct}{24}%
\expandafter\def\csname ok@aoBThreePsZeroPct\endcsname{}%
\newcommand{\aoBThreeRerunHits}{18}%
\expandafter\def\csname ok@aoBThreeRerunHits\endcsname{}%
\newcommand{\aoBThreeRerunUnans}{191}%
\expandafter\def\csname ok@aoBThreeRerunUnans\endcsname{}%
\newcommand{\aoBThreeCtlP}{0.28 and 0.049}%
\expandafter\def\csname ok@aoBThreeCtlP\endcsname{}%
\newcommand{\aoFullSixtyK}{4}%
\expandafter\def\csname ok@aoFullSixtyK\endcsname{}%
\newcommand{\aoFullSixtyExp}{4.6}%
\expandafter\def\csname ok@aoFullSixtyExp\endcsname{}%
\newcommand{\aoFullLongK}{1}%
\expandafter\def\csname ok@aoFullLongK\endcsname{}%
\newcommand{\aoShiftLongK}{none}%
\expandafter\def\csname ok@aoShiftLongK\endcsname{}%
\newcommand{\aoSlackZeroK}{17}%
\expandafter\def\csname ok@aoSlackZeroK\endcsname{}%
\newcommand{\aoSlackZeroN}{27}%
\expandafter\def\csname ok@aoSlackZeroN\endcsname{}%
\newcommand{\aoLocCtlSites}{1}%
\expandafter\def\csname ok@aoLocCtlSites\endcsname{}%
\newcommand{\aoLocCtlP}{0.40}%
\expandafter\def\csname ok@aoLocCtlP\endcsname{}%
\newcommand{\defLocN}{32}%
\expandafter\def\csname ok@defLocN\endcsname{}%
\newcommand{\defLocK}{8}%
\expandafter\def\csname ok@defLocK\endcsname{}%
\newcommand{\defLocEclN}{5}%
\expandafter\def\csname ok@defLocEclN\endcsname{}%
\newcommand{\defLocEclK}{1}%
\expandafter\def\csname ok@defLocEclK\endcsname{}%
\newcommand{\defLocAoN}{27}%
\expandafter\def\csname ok@defLocAoN\endcsname{}%
\newcommand{\defLocAoK}{7}%
\expandafter\def\csname ok@defLocAoK\endcsname{}%
\newcommand{\glNullIneuExpH}{0.17}%
\expandafter\def\csname ok@glNullIneuExpH\endcsname{}%
\newcommand{\glXceuN}{304}%
\expandafter\def\csname ok@glXceuN\endcsname{}%
\newcommand{\glXceuKm}{none}%
\expandafter\def\csname ok@glXceuKm\endcsname{}%
\newcommand{\glNullXceuExpK}{2.39}%
\expandafter\def\csname ok@glNullXceuExpK\endcsname{}%
\newcommand{\confHiK}{12}%
\expandafter\def\csname ok@confHiK\endcsname{}%
\newcommand{\confHiAoK}{10}%
\expandafter\def\csname ok@confHiAoK\endcsname{}%
\newcommand{\confXcyAuc}{0.12}%
\expandafter\def\csname ok@confXcyAuc\endcsname{}%
\newcommand{\eclCleanSites}{19}%
\expandafter\def\csname ok@eclCleanSites\endcsname{}%
\newcommand{\eclCleanSitesHit}{2}%
\expandafter\def\csname ok@eclCleanSitesHit\endcsname{}%
\newcommand{\eclCleanSitesP}{0.006}%
\expandafter\def\csname ok@eclCleanSitesP\endcsname{}%
\newcommand{\eclCleanClipSites}{29}%
\expandafter\def\csname ok@eclCleanClipSites\endcsname{}%
\newcommand{\eclCleanClipSitesHit}{1}%
\expandafter\def\csname ok@eclCleanClipSitesHit\endcsname{}%
\newcommand{\eclCleanClipSitesP}{0.24}%
\expandafter\def\csname ok@eclCleanClipSitesP\endcsname{}%
\newcommand{\auditFalseDecPct}{0.9}%
\expandafter\def\csname ok@auditFalseDecPct\endcsname{}%
\newcommand{\auditConfK}{18}%
\expandafter\def\csname ok@auditConfK\endcsname{}%
\newcommand{\auditConfOneIn}{40}%
\expandafter\def\csname ok@auditConfOneIn\endcsname{}%
\newcommand{\auditCalibXcK}{9}%
\expandafter\def\csname ok@auditCalibXcK\endcsname{}%
\newcommand{\auditRuleV}{20}%
\expandafter\def\csname ok@auditRuleV\endcsname{}%
\newcommand{\auditRuleDlo}{0.6}%
\expandafter\def\csname ok@auditRuleDlo\endcsname{}%
\newcommand{\auditRuleDhi}{1.6}%
\expandafter\def\csname ok@auditRuleDhi\endcsname{}%
\newcommand{\auditRuleZ}{1.5}%
\expandafter\def\csname ok@auditRuleZ\endcsname{}%
\newcommand{\auditRuleW}{8}%
\expandafter\def\csname ok@auditRuleW\endcsname{}%
\newcommand{\auditRuleGap}{2}%
\expandafter\def\csname ok@auditRuleGap\endcsname{}%
\newcommand{\auditFreezeLagMin}{36}%
\expandafter\def\csname ok@auditFreezeLagMin\endcsname{}%
\newcommand{\auditXcConfOneAc}{17}%
\expandafter\def\csname ok@auditXcConfOneAc\endcsname{}%
\newcommand{\auditXcConfMinK}{12}%
\expandafter\def\csname ok@auditXcConfMinK\endcsname{}%
\newcommand{\auditInjDecRec}{7}%
\expandafter\def\csname ok@auditInjDecRec\endcsname{}%
\newcommand{\auditXcConfRec}{7}%
\expandafter\def\csname ok@auditXcConfRec\endcsname{}%
\newcommand{\auditXcConfTopTwo}{12}%
\expandafter\def\csname ok@auditXcConfTopTwo\endcsname{}%
\newcommand{\liveSlackS}{30}%
\expandafter\def\csname ok@liveSlackS\endcsname{}%
\newcommand{\liveLbBoxKm}{222}%
\expandafter\def\csname ok@liveLbBoxKm\endcsname{}%
\newcommand{\liveLtBoxKm}{22}%
\expandafter\def\csname ok@liveLtBoxKm\endcsname{}%
\newcommand{\liveSamplePct}{12.5}%
\expandafter\def\csname ok@liveSamplePct\endcsname{}%
\newcommand{\liveLtK}{3}%
\expandafter\def\csname ok@liveLtK\endcsname{}%
\newcommand{\liveLtN}{143}%
\expandafter\def\csname ok@liveLtN\endcsname{}%
\newcommand{\liveLtCzK}{1}%
\expandafter\def\csname ok@liveLtCzK\endcsname{}%
\newcommand{\liveLtCzN}{79}%
\expandafter\def\csname ok@liveLtCzN\endcsname{}%
\newcommand{\liveScStreams}{7}%
\expandafter\def\csname ok@liveScStreams\endcsname{}%
\newcommand{\liveWindows}{147}%
\expandafter\def\csname ok@liveWindows\endcsname{}%
\newcommand{\liveExp}{0.009}%
\expandafter\def\csname ok@liveExp\endcsname{}%
\newcommand{\liveCzN}{80}%
\expandafter\def\csname ok@liveCzN\endcsname{}%
\newcommand{\liveK}{1}%
\expandafter\def\csname ok@liveK\endcsname{}%
\newcommand{\scRecordings}{11{,}698{,}054}%
\expandafter\def\csname ok@scRecordings\endcsname{}%
\newcommand{\scRecPass}{1{,}956{,}202}%
\expandafter\def\csname ok@scRecPass\endcsname{}%
\newcommand{\scListed}{11{,}727{,}009}%
\expandafter\def\csname ok@scListed\endcsname{}%
\newcommand{\scListedDays}{1{,}359}%
\expandafter\def\csname ok@scListedDays\endcsname{}%
\newcommand{\scRecM}{11.7~million}%
\expandafter\def\csname ok@scRecM\endcsname{}%
\newcommand{\scRecPassM}{2~million}%
\expandafter\def\csname ok@scRecPassM\endcsname{}%
\newcommand{\scAssumedWin}{1{,}245{,}429}%
\expandafter\def\csname ok@scAssumedWin\endcsname{}%
\newcommand{\scPasses}{5{,}243{,}380}%
\expandafter\def\csname ok@scPasses\endcsname{}%
\newcommand{\scSources}{21}%
\expandafter\def\csname ok@scSources\endcsname{}%
\newcommand{\scDays}{1{,}356}%
\expandafter\def\csname ok@scDays\endcsname{}%
\newcommand{\scNoCov}{1{,}805{,}595}%
\expandafter\def\csname ok@scNoCov\endcsname{}%
\newcommand{\scPublic}{3{,}912{,}415}%
\expandafter\def\csname ok@scPublic\endcsname{}%
\newcommand{\scPubRecPass}{581{,}426}%
\expandafter\def\csname ok@scPubRecPass\endcsname{}%
\newcommand{\scPubPasses}{2{,}218{,}999}%
\expandafter\def\csname ok@scPubPasses\endcsname{}%
\newcommand{\scPubPassPct}{14.9}%
\expandafter\def\csname ok@scPubPassPct\endcsname{}%
\newcommand{\scXcN}{9{,}579}%
\expandafter\def\csname ok@scXcN\endcsname{}%
\newcommand{\scXcSame}{9{,}250}%
\expandafter\def\csname ok@scXcSame\endcsname{}%
\newcommand{\dsRecordings}{11{,}647}%
\expandafter\def\csname ok@dsRecordings\endcsname{}%
\newcommand{\dsSources}{6}%
\expandafter\def\csname ok@dsSources\endcsname{}%
\newcommand{\dsPasses}{4{,}020}%
\expandafter\def\csname ok@dsPasses\endcsname{}%
\newcommand{\dsRecPass}{2{,}137}%
\expandafter\def\csname ok@dsRecPass\endcsname{}%
\newcommand{\dsRadiusKm}{10}%
\expandafter\def\csname ok@dsRadiusKm\endcsname{}%
\newcommand{\dsName}{AircraftWitness}%
\expandafter\def\csname ok@dsName\endcsname{}%
\newcommand{\dsDetRec}{729}%
\expandafter\def\csname ok@dsDetRec\endcsname{}%
\newcommand{\dsDetNearK}{25}%
\expandafter\def\csname ok@dsDetNearK\endcsname{}%
\newcommand{\dsDetNearN}{48}%
\expandafter\def\csname ok@dsDetNearN\endcsname{}%
\newcommand{\dsDetNearHi}{2}%
\expandafter\def\csname ok@dsDetNearHi\endcsname{}%
\newcommand{\dsDetFarK}{122}%
\expandafter\def\csname ok@dsDetFarK\endcsname{}%
\newcommand{\dsDetFarN}{631}%
\expandafter\def\csname ok@dsDetFarN\endcsname{}%
\newcommand{\dsDetFarLo}{5}%
\expandafter\def\csname ok@dsDetFarLo\endcsname{}%
\newcommand{\dsDetFarHi}{10}%
\expandafter\def\csname ok@dsDetFarHi\endcsname{}%
\newcommand{\dsDetUnmatchedHr}{43}%
\expandafter\def\csname ok@dsDetUnmatchedHr\endcsname{}%
\newcommand{\unhAoK}{5}%
\expandafter\def\csname ok@unhAoK\endcsname{}%
\newcommand{\unhAoN}{295}%
\expandafter\def\csname ok@unhAoN\endcsname{}%
\newcommand{\zenSites}{48}%
\expandafter\def\csname ok@zenSites\endcsname{}%
\newcommand{\zenUnbuilt}{3}%
\expandafter\def\csname ok@zenUnbuilt\endcsname{}%
\newcommand{\zenLtK}{none}%
\expandafter\def\csname ok@zenLtK\endcsname{}%
\newcommand{\zenLtN}{229}%
\expandafter\def\csname ok@zenLtN\endcsname{}%
\newcommand{\zenN}{232}%
\expandafter\def\csname ok@zenN\endcsname{}%
\newcommand{\zenCtlK}{none}%
\expandafter\def\csname ok@zenCtlK\endcsname{}%
\newcommand{\zenCtlN}{72}%
\expandafter\def\csname ok@zenCtlN\endcsname{}%
\newcommand{\audN}{2{,}000}%
\expandafter\def\csname ok@audN\endcsname{}%
\newcommand{\audPct}{66}%
\expandafter\def\csname ok@audPct\endcsname{}%
\newcommand{\audDecoyPct}{23}%
\expandafter\def\csname ok@audDecoyPct\endcsname{}%
\newcommand{\audNearPct}{81}%
\expandafter\def\csname ok@audNearPct\endcsname{}%
\newcommand{\audFarPct}{49}%
\expandafter\def\csname ok@audFarPct\endcsname{}%
\newcommand{\audMidPct}{88}%
\expandafter\def\csname ok@audMidPct\endcsname{}%
\newcommand{\audCi}{52 to 78\%}%
\expandafter\def\csname ok@audCi\endcsname{}%
\newcommand{\audDet}{825}%
\expandafter\def\csname ok@audDet\endcsname{}%
\newcommand{\audClfPct}{9.8}%
\expandafter\def\csname ok@audClfPct\endcsname{}%
\newcommand{\audClfDecoyPct}{1.1}%
\expandafter\def\csname ok@audClfDecoyPct\endcsname{}%
\newcommand{\audUnwPct}{42}%
\expandafter\def\csname ok@audUnwPct\endcsname{}%
\newcommand{\audAgreePct}{64}%
\expandafter\def\csname ok@audAgreePct\endcsname{}%
\newcommand{\audKappa}{0.15}%
\expandafter\def\csname ok@audKappa\endcsname{}%
\newcommand{\rrUnhK}{12}%
\expandafter\def\csname ok@rrUnhK\endcsname{}%
\newcommand{\rrUnhCtlMax}{1}%
\expandafter\def\csname ok@rrUnhCtlMax\endcsname{}%
\newcommand{\phThreePlN}{7}%
\expandafter\def\csname ok@phThreePlN\endcsname{}%
\newcommand{\phThreeMoN}{252}%
\expandafter\def\csname ok@phThreeMoN\endcsname{}%
\newcommand{\phThreeRefPlK}{4}%
\expandafter\def\csname ok@phThreeRefPlK\endcsname{}%
\newcommand{\phThreeRefPlP}{0.16}%
\expandafter\def\csname ok@phThreeRefPlP\endcsname{}%
\newcommand{\phThreePlExp}{2.3}%
\expandafter\def\csname ok@phThreePlExp\endcsname{}%
\newcommand{\phThreeRefMoK}{17}%
\expandafter\def\csname ok@phThreeRefMoK\endcsname{}%
\newcommand{\phThreeRefMoP}{$5\times10^{-9}$}%
\expandafter\def\csname ok@phThreeRefMoP\endcsname{}%
\newcommand{\phThreeMoExp}{2.8}%
\expandafter\def\csname ok@phThreeMoExp\endcsname{}%
\newcommand{\phThreeDpPlK}{5}%
\expandafter\def\csname ok@phThreeDpPlK\endcsname{}%
\newcommand{\phThreeDpPlP}{0.039}%
\expandafter\def\csname ok@phThreeDpPlP\endcsname{}%
\newcommand{\phThreeDpMoK}{23}%
\expandafter\def\csname ok@phThreeDpMoK\endcsname{}%
\newcommand{\phThreeDpMoP}{$2\times10^{-14}$}%
\expandafter\def\csname ok@phThreeDpMoP\endcsname{}%
\newcommand{\phThreeUnhPlK}{4}%
\expandafter\def\csname ok@phThreeUnhPlK\endcsname{}%
\newcommand{\phThreeUnhPlP}{0.16}%
\expandafter\def\csname ok@phThreeUnhPlP\endcsname{}%
\newcommand{\phThreeUnhMoK}{10}%
\expandafter\def\csname ok@phThreeUnhMoK\endcsname{}%
\newcommand{\phThreeUnhMoP}{$6\times10^{-4}$}%
\expandafter\def\csname ok@phThreeUnhMoP\endcsname{}%
\newcommand{\phFourPlN}{3}%
\expandafter\def\csname ok@phFourPlN\endcsname{}%
\newcommand{\phFourMoN}{86}%
\expandafter\def\csname ok@phFourMoN\endcsname{}%
\newcommand{\phFourPlUsed}{2}%
\expandafter\def\csname ok@phFourPlUsed\endcsname{}%
\newcommand{\phFourRefPlK}{2}%
\expandafter\def\csname ok@phFourRefPlK\endcsname{}%
\newcommand{\phFourRefPlP}{0.12}%
\expandafter\def\csname ok@phFourRefPlP\endcsname{}%
\newcommand{\phFourRefMoK}{4}%
\expandafter\def\csname ok@phFourRefMoK\endcsname{}%
\newcommand{\phFourRefMoP}{0.0078}%
\expandafter\def\csname ok@phFourRefMoP\endcsname{}%
\newcommand{\bThreePlCiRef}{2 to 9\%}%
\expandafter\def\csname ok@bThreePlCiRef\endcsname{}%
\newcommand{\bThreePlCiDp}{3 to 14\%}%
\expandafter\def\csname ok@bThreePlCiDp\endcsname{}%
\newcommand{\swRefOrigK}{21}%
\expandafter\def\csname ok@swRefOrigK\endcsname{}%
\newcommand{\swRefOrigAns}{98}%
\expandafter\def\csname ok@swRefOrigAns\endcsname{}%
\newcommand{\swRefOrigMoK}{17}%
\expandafter\def\csname ok@swRefOrigMoK\endcsname{}%
\newcommand{\swRefOrigMoP}{$5\times10^{-9}$}%
\expandafter\def\csname ok@swRefOrigMoP\endcsname{}%
\newcommand{\swOrigExp}{2.8}%
\expandafter\def\csname ok@swOrigExp\endcsname{}%
\newcommand{\swDpOrigK}{29}%
\expandafter\def\csname ok@swDpOrigK\endcsname{}%
\newcommand{\swDpOrigAns}{157}%
\expandafter\def\csname ok@swDpOrigAns\endcsname{}%
\newcommand{\swDpOrigMoK}{23}%
\expandafter\def\csname ok@swDpOrigMoK\endcsname{}%
\newcommand{\swDpOrigMoP}{$2\times10^{-14}$}%
\expandafter\def\csname ok@swDpOrigMoP\endcsname{}%
\newcommand{\swRefSlFiveK}{36}%
\expandafter\def\csname ok@swRefSlFiveK\endcsname{}%
\newcommand{\swRefSlFiveAns}{86}%
\expandafter\def\csname ok@swRefSlFiveAns\endcsname{}%
\newcommand{\swRefSlFiveMoK}{26}%
\expandafter\def\csname ok@swRefSlFiveMoK\endcsname{}%
\newcommand{\swRefSlFiveMoP}{$2\times10^{-17}$}%
\expandafter\def\csname ok@swRefSlFiveMoP\endcsname{}%
\newcommand{\swSlFiveExp}{2.8}%
\expandafter\def\csname ok@swSlFiveExp\endcsname{}%
\newcommand{\swDpSlFiveK}{41}%
\expandafter\def\csname ok@swDpSlFiveK\endcsname{}%
\newcommand{\swDpSlFiveAns}{147}%
\expandafter\def\csname ok@swDpSlFiveAns\endcsname{}%
\newcommand{\swDpSlFiveMoK}{30}%
\expandafter\def\csname ok@swDpSlFiveMoK\endcsname{}%
\newcommand{\swDpSlFiveMoP}{$1\times10^{-21}$}%
\expandafter\def\csname ok@swDpSlFiveMoP\endcsname{}%
\newcommand{\swRefSlLongK}{17}%
\expandafter\def\csname ok@swRefSlLongK\endcsname{}%
\newcommand{\swRefSlLongAns}{130}%
\expandafter\def\csname ok@swRefSlLongAns\endcsname{}%
\newcommand{\swRefSlLongMoK}{15}%
\expandafter\def\csname ok@swRefSlLongMoK\endcsname{}%
\newcommand{\swRefSlLongMoP}{$2\times10^{-7}$}%
\expandafter\def\csname ok@swRefSlLongMoP\endcsname{}%
\newcommand{\swRefSlLongPlK}{4}%
\expandafter\def\csname ok@swRefSlLongPlK\endcsname{}%
\newcommand{\swSlLongExp}{2.8}%
\expandafter\def\csname ok@swSlLongExp\endcsname{}%
\newcommand{\swDpSlLongK}{18}%
\expandafter\def\csname ok@swDpSlLongK\endcsname{}%
\newcommand{\swDpSlLongAns}{177}%
\expandafter\def\csname ok@swDpSlLongAns\endcsname{}%
\newcommand{\swDpSlLongMoK}{15}%
\expandafter\def\csname ok@swDpSlLongMoK\endcsname{}%
\newcommand{\swDpSlLongMoP}{$2\times10^{-7}$}%
\expandafter\def\csname ok@swDpSlLongMoP\endcsname{}%
\newcommand{\swDpSlLongPlK}{4}%
\expandafter\def\csname ok@swDpSlLongPlK\endcsname{}%
\newcommand{\swDpLongTo}{3}%
\expandafter\def\csname ok@swDpLongTo\endcsname{}%
\newcommand{\swRefSqTenK}{28}%
\expandafter\def\csname ok@swRefSqTenK\endcsname{}%
\newcommand{\swRefSqTenAns}{80}%
\expandafter\def\csname ok@swRefSqTenAns\endcsname{}%
\newcommand{\swRefSqTenMoK}{23}%
\expandafter\def\csname ok@swRefSqTenMoK\endcsname{}%
\newcommand{\swRefSqTenMoP}{$9\times10^{-4}$}%
\expandafter\def\csname ok@swRefSqTenMoP\endcsname{}%
\newcommand{\swSqTenExp}{11.3}%
\expandafter\def\csname ok@swSqTenExp\endcsname{}%
\newcommand{\swDpSqTenK}{32}%
\expandafter\def\csname ok@swDpSqTenK\endcsname{}%
\newcommand{\swDpSqTenAns}{133}%
\expandafter\def\csname ok@swDpSqTenAns\endcsname{}%
\newcommand{\swDpSqTenMoK}{25}%
\expandafter\def\csname ok@swDpSqTenMoK\endcsname{}%
\newcommand{\swDpSqTenMoP}{$2\times10^{-4}$}%
\expandafter\def\csname ok@swDpSqTenMoP\endcsname{}%
\newcommand{\swRefSqFiveK}{29}%
\expandafter\def\csname ok@swRefSqFiveK\endcsname{}%
\newcommand{\swRefSqFiveAns}{70}%
\expandafter\def\csname ok@swRefSqFiveAns\endcsname{}%
\newcommand{\swRefSqFiveMoK}{22}%
\expandafter\def\csname ok@swRefSqFiveMoK\endcsname{}%
\newcommand{\swRefSqFiveMoP}{$>$0.99}%
\expandafter\def\csname ok@swRefSqFiveMoP\endcsname{}%
\newcommand{\swSqFiveExp}{45.2}%
\expandafter\def\csname ok@swSqFiveExp\endcsname{}%
\newcommand{\swN}{360}%
\expandafter\def\csname ok@swN\endcsname{}%
\newcommand{\swDpSqFiveK}{38}%
\expandafter\def\csname ok@swDpSqFiveK\endcsname{}%
\newcommand{\swDpSqFiveAns}{117}%
\expandafter\def\csname ok@swDpSqFiveAns\endcsname{}%
\newcommand{\swDpSqFiveMoK}{30}%
\expandafter\def\csname ok@swDpSqFiveMoK\endcsname{}%
\newcommand{\swDpSqFiveMoP}{$>$0.99}%
\expandafter\def\csname ok@swDpSqFiveMoP\endcsname{}%
\newcommand{\prevOosN}{1{,}556}%
\expandafter\def\csname ok@prevOosN\endcsname{}%
\newcommand{\prevOosRefPct}{3.0}%
\expandafter\def\csname ok@prevOosRefPct\endcsname{}%
\newcommand{\prevOosRefK}{46}%
\expandafter\def\csname ok@prevOosRefK\endcsname{}%
\newcommand{\prevOosRefCi}{1.9 to 4.3\%}%
\expandafter\def\csname ok@prevOosRefCi\endcsname{}%
\newcommand{\prevOosUnhPct}{1.7}%
\expandafter\def\csname ok@prevOosUnhPct\endcsname{}%
\newcommand{\prevOosUnhCi}{1.0 to 2.5\%}%
\expandafter\def\csname ok@prevOosUnhCi\endcsname{}%
\newcommand{\prevOosEcoRefK}{39}%
\expandafter\def\csname ok@prevOosEcoRefK\endcsname{}%
\newcommand{\prevOosRestRefK}{7}%
\expandafter\def\csname ok@prevOosRestRefK\endcsname{}%
\newcommand{\prevOosRestN}{1{,}069}%
\expandafter\def\csname ok@prevOosRestN\endcsname{}%
\newcommand{\prevOosRestRefPct}{0.65}%
\expandafter\def\csname ok@prevOosRestRefPct\endcsname{}%
\newcommand{\prevTopDevKm}{0.3}%
\expandafter\def\csname ok@prevTopDevKm\endcsname{}%
\newcommand{\hmBThreeJetN}{120}%
\expandafter\def\csname ok@hmBThreeJetN\endcsname{}%
\newcommand{\hmBThreeFarN}{197}%
\expandafter\def\csname ok@hmBThreeFarN\endcsname{}%
\newcommand{\hmBThreePropK}{19}%
\expandafter\def\csname ok@hmBThreePropK\endcsname{}%
\newcommand{\hmBThreePropN}{187}%
\expandafter\def\csname ok@hmBThreePropN\endcsname{}%
\newcommand{\hmBThreeRotK}{2}%
\expandafter\def\csname ok@hmBThreeRotK\endcsname{}%
\newcommand{\hmBThreeRotN}{22}%
\expandafter\def\csname ok@hmBThreeRotN\endcsname{}%
\newcommand{\hmBThreeNoAc}{23}%
\expandafter\def\csname ok@hmBThreeNoAc\endcsname{}%
\newcommand{\hmPrNoDetN}{1{,}064}%
\expandafter\def\csname ok@hmPrNoDetN\endcsname{}%
\newcommand{\lvTwoN}{686}%
\expandafter\def\csname ok@lvTwoN\endcsname{}%
\newcommand{\lvTwoStreams}{39}%
\expandafter\def\csname ok@lvTwoStreams\endcsname{}%
\newcommand{\lvTwoExp}{5.39}%
\expandafter\def\csname ok@lvTwoExp\endcsname{}%
\newcommand{\lvTwoRefK}{7}%
\expandafter\def\csname ok@lvTwoRefK\endcsname{}%
\newcommand{\lvTwoRefSt}{3}%
\expandafter\def\csname ok@lvTwoRefSt\endcsname{}%
\newcommand{\lvTwoRefStP}{0.015}%
\expandafter\def\csname ok@lvTwoRefStP\endcsname{}%
\newcommand{\lvTwoRefUniP}{0.30}%
\expandafter\def\csname ok@lvTwoRefUniP\endcsname{}%
\newcommand{\lvTwoRefTopK}{5}%
\expandafter\def\csname ok@lvTwoRefTopK\endcsname{}%
\newcommand{\lvTwoRefStUniP}{0.90}%
\expandafter\def\csname ok@lvTwoRefStUniP\endcsname{}%
\newcommand{\lvTwoDpK}{2}%
\expandafter\def\csname ok@lvTwoDpK\endcsname{}%
\newcommand{\lvTwoDpSt}{2}%
\expandafter\def\csname ok@lvTwoDpSt\endcsname{}%
\newcommand{\lvTwoDpStP}{0.007}%
\expandafter\def\csname ok@lvTwoDpStP\endcsname{}%
\newcommand{\lvTwoDpUniP}{0.97}%
\expandafter\def\csname ok@lvTwoDpUniP\endcsname{}%
\newcommand{\lvTwoDpStUniP}{0.97}%
\expandafter\def\csname ok@lvTwoDpStUniP\endcsname{}%
\newcommand{\lvTwoUnhK}{4}%
\expandafter\def\csname ok@lvTwoUnhK\endcsname{}%
\newcommand{\lvTwoUnhSt}{2}%
\expandafter\def\csname ok@lvTwoUnhSt\endcsname{}%
\newcommand{\lvTwoUnhStP}{0.049}%
\expandafter\def\csname ok@lvTwoUnhStP\endcsname{}%
\newcommand{\lvTwoUnhUniP}{0.79}%
\expandafter\def\csname ok@lvTwoUnhUniP\endcsname{}%
\newcommand{\lvTwoUnhStUniP}{0.97}%
\expandafter\def\csname ok@lvTwoUnhStUniP\endcsname{}%
\newcommand{\lvTwoClkRef}{38}%
\expandafter\def\csname ok@lvTwoClkRef\endcsname{}%
\newcommand{\lvTwoClkNull}{15}%
\expandafter\def\csname ok@lvTwoClkNull\endcsname{}%
\newcommand{\lvTwoClkDp}{38}%
\expandafter\def\csname ok@lvTwoClkDp\endcsname{}%
\newcommand{\lvTwoWtN}{184}%
\expandafter\def\csname ok@lvTwoWtN\endcsname{}%
\newcommand{\lvTwoWaN}{151}%
\expandafter\def\csname ok@lvTwoWaN\endcsname{}%
\newcommand{\lvWcSt}{9}%
\expandafter\def\csname ok@lvWcSt\endcsname{}%
\newcommand{\lvWcWin}{110}%
\expandafter\def\csname ok@lvWcWin\endcsname{}%
\newcommand{\lvWcCtl}{157}%
\expandafter\def\csname ok@lvWcCtl\endcsname{}%
\newcommand{\lvTwoLsSt}{30}%
\expandafter\def\csname ok@lvTwoLsSt\endcsname{}%
\newcommand{\lvTwoApSt}{3}%
\expandafter\def\csname ok@lvTwoApSt\endcsname{}%
\newcommand{\lvTwoCfSt}{2}%
\expandafter\def\csname ok@lvTwoCfSt\endcsname{}%
\newcommand{\lvTwoIcSt}{4}%
\expandafter\def\csname ok@lvTwoIcSt\endcsname{}%
\newcommand{\lvRegN}{796}%
\expandafter\def\csname ok@lvRegN\endcsname{}%
\newcommand{\lvRegStreams}{48}%
\expandafter\def\csname ok@lvRegStreams\endcsname{}%
\newcommand{\lvRegRefStP}{0.020}%
\expandafter\def\csname ok@lvRegRefStP\endcsname{}%
\newcommand{\lvRegRefStPBonf}{0.059}%
\expandafter\def\csname ok@lvRegRefStPBonf\endcsname{}%
\newcommand{\lvRegDpStP}{0.016}%
\expandafter\def\csname ok@lvRegDpStP\endcsname{}%
\newcommand{\lvRegDpStPBonf}{0.047}%
\expandafter\def\csname ok@lvRegDpStPBonf\endcsname{}%
\newcommand{\lvRegUnhStP}{0.060}%
\expandafter\def\csname ok@lvRegUnhStP\endcsname{}%
\newcommand{\lvRegUnhStPBonf}{0.18}%
\expandafter\def\csname ok@lvRegUnhStPBonf\endcsname{}%
\newcommand{\lvRegStallN}{10}%
\expandafter\def\csname ok@lvRegStallN\endcsname{}%
\newcommand{\lvRegBands}{7}%
\expandafter\def\csname ok@lvRegBands\endcsname{}%
\newcommand{\bFourN}{99}%
\expandafter\def\csname ok@bFourN\endcsname{}%
\newcommand{\bFourSites}{6}%
\expandafter\def\csname ok@bFourSites\endcsname{}%
\newcommand{\bFourSitesExp}{0.7}%
\expandafter\def\csname ok@bFourSitesExp\endcsname{}%
\newcommand{\bFourRefSitesHit}{4}%
\expandafter\def\csname ok@bFourRefSitesHit\endcsname{}%
\newcommand{\bFourRefP}{0.0026}%
\expandafter\def\csname ok@bFourRefP\endcsname{}%
\newcommand{\bFourRefPLoo}{0.014}%
\expandafter\def\csname ok@bFourRefPLoo\endcsname{}%
\newcommand{\bFourDpP}{0.16}%
\expandafter\def\csname ok@bFourDpP\endcsname{}%
\newcommand{\bFourUnhP}{0.54}%
\expandafter\def\csname ok@bFourUnhP\endcsname{}%
\newcommand{\jbMaxM}{279}%
\expandafter\def\csname ok@jbMaxM\endcsname{}%
\newcommand{\jbTotM}{397}%
\expandafter\def\csname ok@jbTotM\endcsname{}%
\newcommand{\jbTotOffM}{367}%
\expandafter\def\csname ok@jbTotOffM\endcsname{}%
\newcommand{\jbTotGpsM}{30}%
\expandafter\def\csname ok@jbTotGpsM\endcsname{}%
\newcommand{\jbTotRefWorstK}{17}%
\expandafter\def\csname ok@jbTotRefWorstK\endcsname{}%
\newcommand{\jbTotDpWorstK}{17}%
\expandafter\def\csname ok@jbTotDpWorstK\endcsname{}%
\newcommand{\jbGpsKm}{1.8}%
\expandafter\def\csname ok@jbGpsKm\endcsname{}%
\newcommand{\prevN}{3{,}000}%
\expandafter\def\csname ok@prevN\endcsname{}%
\newcommand{\prevSnapPassPct}{6.8}%
\expandafter\def\csname ok@prevSnapPassPct\endcsname{}%
\newcommand{\prevFunOos}{0.10}%
\expandafter\def\csname ok@prevFunOos\endcsname{}%
\newcommand{\prevExpPct}{0.79}%
\expandafter\def\csname ok@prevExpPct\endcsname{}%
\newcommand{\prevDpPct}{0.17}%
\expandafter\def\csname ok@prevDpPct\endcsname{}%
\newcommand{\prevUnhPct}{5.7}%
\expandafter\def\csname ok@prevUnhPct\endcsname{}%
\newcommand{\prevRefPct}{9.5}%
\expandafter\def\csname ok@prevRefPct\endcsname{}%
\newcommand{\prevRefAoK}{238}%
\expandafter\def\csname ok@prevRefAoK\endcsname{}%
\newcommand{\prevAoNs}{1{,}232}%
\expandafter\def\csname ok@prevAoNs\endcsname{}%
\newcommand{\prevRefXcN}{221}%
\expandafter\def\csname ok@prevRefXcN\endcsname{}%
\newcommand{\prevRefNoAoPct}{2.7}%
\expandafter\def\csname ok@prevRefNoAoPct\endcsname{}%
\newcommand{\prevRefCiPl}{4.6 to 14.2\%}%
\expandafter\def\csname ok@prevRefCiPl\endcsname{}%
\newcommand{\prevRefK}{285}%
\expandafter\def\csname ok@prevRefK\endcsname{}%
\newcommand{\prevRefTopK}{103}%
\expandafter\def\csname ok@prevRefTopK\endcsname{}%
\newcommand{\prevRefNoTop}{6.9}%
\expandafter\def\csname ok@prevRefNoTop\endcsname{}%
\newcommand{\prevUnhCiPl}{2.7 to 8.3\%}%
\expandafter\def\csname ok@prevUnhCiPl\endcsname{}%
\newcommand{\prevUnhK}{171}%
\expandafter\def\csname ok@prevUnhK\endcsname{}%
\newcommand{\prevTopK}{55}%
\expandafter\def\csname ok@prevTopK\endcsname{}%
\newcommand{\prevUnhNoTop}{4.4}%
\expandafter\def\csname ok@prevUnhNoTop\endcsname{}%
\newcommand{\prevWtPct}{0.17}%
\expandafter\def\csname ok@prevWtPct\endcsname{}%
\newcommand{\prevWaPct}{0.33}%
\expandafter\def\csname ok@prevWaPct\endcsname{}%
\newcommand{\anEdgeSitesW}{1 recorder}%
\expandafter\def\csname ok@anEdgeSitesW\endcsname{}%
\newcommand{\anEdgeStSites}{13}%
\expandafter\def\csname ok@anEdgeStSites\endcsname{}%
\newcommand{\anDetEqRecall}{28}%
\expandafter\def\csname ok@anDetEqRecall\endcsname{}%
\newcommand{\anDetEqDecoy}{22}%
\expandafter\def\csname ok@anDetEqDecoy\endcsname{}%
\newcommand{\anDetEqP}{0.011}%
\expandafter\def\csname ok@anDetEqP\endcsname{}%
\newcommand{\anDetEqPairs}{1{,}314}%
\expandafter\def\csname ok@anDetEqPairs\endcsname{}%
\newcommand{\anDetEqRec}{270}%
\expandafter\def\csname ok@anDetEqRec\endcsname{}%
\newcommand{\aoBThreeSitesExp}{2.7}%
\expandafter\def\csname ok@aoBThreeSitesExp\endcsname{}%
\newcommand{\aoBThreeP}{$2\times10^{-7}$}%
\expandafter\def\csname ok@aoBThreeP\endcsname{}%
\newcommand{\aoBThreeTwoLookP}{$4\times10^{-7}$}%
\expandafter\def\csname ok@aoBThreeTwoLookP\endcsname{}%
\newcommand{\unhBThreeP}{0.0057}%
\expandafter\def\csname ok@unhBThreeP\endcsname{}%
\newcommand{\unhAoP}{$5\times10^{-5}$}%
\expandafter\def\csname ok@unhAoP\endcsname{}%
\newcommand{\aoLocSitesP}{0.044}%
\expandafter\def\csname ok@aoLocSitesP\endcsname{}%
\newcommand{\eclSitesP}{0.17}%
\expandafter\def\csname ok@eclSitesP\endcsname{}%
\newcommand{\eclPoolSitesP}{0.006}%
\expandafter\def\csname ok@eclPoolSitesP\endcsname{}%
\newcommand{\anAuc}{0.82}%
\expandafter\def\csname ok@anAuc\endcsname{}%
\newcommand{\anAucN}{98}%
\expandafter\def\csname ok@anAucN\endcsname{}%
\newcommand{\anLoo}{17 to 18}%
\expandafter\def\csname ok@anLoo\endcsname{}%
\newcommand{\anYieldRefCi}{3 to 8\%}%
\expandafter\def\csname ok@anYieldRefCi\endcsname{}%
\newcommand{\anYieldDpCi}{5 to 12\%}%
\expandafter\def\csname ok@anYieldDpCi\endcsname{}%
\newcommand{\anPopPerLo}{6}%
\expandafter\def\csname ok@anPopPerLo\endcsname{}%
\newcommand{\anPopPerHi}{18}%
\expandafter\def\csname ok@anPopPerHi\endcsname{}%
\newcommand{\anPopSensors}{23}%
\expandafter\def\csname ok@anPopSensors\endcsname{}%
\newcommand{\eeRefK}{2}%
\expandafter\def\csname ok@eeRefK\endcsname{}%
\newcommand{\eeRefCi}{0.00 to 0.31\%}%
\expandafter\def\csname ok@eeRefCi\endcsname{}%
\newcommand{\eeDpK}{0}%
\expandafter\def\csname ok@eeDpK\endcsname{}%
\newcommand{\eeUnhK}{1}%
\expandafter\def\csname ok@eeUnhK\endcsname{}%
\newcommand{\eeExpPct}{0.79}%
\expandafter\def\csname ok@eeExpPct\endcsname{}%
\newcommand{\eeRefAns}{13}%
\expandafter\def\csname ok@eeRefAns\endcsname{}%
\newcommand{\eeDpAns}{7}%
\expandafter\def\csname ok@eeDpAns\endcsname{}%
\newcommand{\eeUnhAns}{20}%
\expandafter\def\csname ok@eeUnhAns\endcsname{}%
\newcommand{\eeAnsMax}{20}%
\expandafter\def\csname ok@eeAnsMax\endcsname{}%
\newcommand{\eeHitMax}{2}%
\expandafter\def\csname ok@eeHitMax\endcsname{}%
\newcommand{\eeCtlMaxK}{1}%
\expandafter\def\csname ok@eeCtlMaxK\endcsname{}%
\newcommand{\eeReady}{1{,}378}%
\expandafter\def\csname ok@eeReady\endcsname{}%
\newcommand{\eeUnrun}{73}%
\expandafter\def\csname ok@eeUnrun\endcsname{}%
\newcommand{\eeN}{1{,}600}%
\expandafter\def\csname ok@eeN\endcsname{}%
\newcommand{\frRefK}{2}%
\expandafter\def\csname ok@frRefK\endcsname{}%
\newcommand{\frRefExp}{0.20}%
\expandafter\def\csname ok@frRefExp\endcsname{}%
\newcommand{\frRefP}{0.016}%
\expandafter\def\csname ok@frRefP\endcsname{}%
\newcommand{\frRefAns}{30}%
\expandafter\def\csname ok@frRefAns\endcsname{}%
\newcommand{\frDpK}{4}%
\expandafter\def\csname ok@frDpK\endcsname{}%
\newcommand{\frDpExp}{0.10}%
\expandafter\def\csname ok@frDpExp\endcsname{}%
\newcommand{\frDpP}{$7\times10^{-7}$}%
\expandafter\def\csname ok@frDpP\endcsname{}%
\newcommand{\frDpPv}{0.000001}%
\expandafter\def\csname ok@frDpPv\endcsname{}%
\newcommand{\frDpHits}{4}%
\expandafter\def\csname ok@frDpHits\endcsname{}%
\newcommand{\frDpAns}{16}%
\expandafter\def\csname ok@frDpAns\endcsname{}%
\newcommand{\frUnhK}{1}%
\expandafter\def\csname ok@frUnhK\endcsname{}%
\newcommand{\frUnhExp}{0.15}%
\expandafter\def\csname ok@frUnhExp\endcsname{}%
\newcommand{\frUnhP}{0.14}%
\expandafter\def\csname ok@frUnhP\endcsname{}%
\newcommand{\frUnhPv}{0.138338}%
\expandafter\def\csname ok@frUnhPv\endcsname{}%
\newcommand{\frUnhAns}{15}%
\expandafter\def\csname ok@frUnhAns\endcsname{}%
\newcommand{\frExpU}{0.57}%
\expandafter\def\csname ok@frExpU\endcsname{}%
\newcommand{\frCtlMaxK}{1}%
\expandafter\def\csname ok@frCtlMaxK\endcsname{}%
\newcommand{\frBestPc}{0.0065}%
\expandafter\def\csname ok@frBestPc\endcsname{}%
\expandafter\def\csname ok@frOK\endcsname{}%
\newcommand{\frClips}{73}%
\expandafter\def\csname ok@frClips\endcsname{}%
\newcommand{\frN}{24}%
\expandafter\def\csname ok@frN\endcsname{}%

%% file: build/figready.tex
\expandafter\def\csname figok@central\endcsname{}%
\expandafter\def\csname figok@clock\endcsname{}%
\expandafter\def\csname figok@clocklv\endcsname{}%
\expandafter\def\csname figok@dataset\endcsname{}%
\expandafter\def\csname figok@doppler\endcsname{}%
\expandafter\def\csname figok@fieldtests\endcsname{}%
\expandafter\def\csname figok@fieldtests2\endcsname{}%
\expandafter\def\csname figok@worth\endcsname{}%

%% file: build/authors.tex
\author{%
  Kris Atallah \\
  New York University \\
  New York, NY, USA \\
  \texttt{kris.a@nyu.edu}}

%% file: sec/abstract.tex
\begin{abstract}
Aircraft overhead can reveal where an outdoor recording was made. Propellers and rotors leave Doppler curves in audio, and public ADS-B archives log equipped aircraft. We measure how often, blind and against chance. Our main test was registered and timestamped before any audio, and its two amendments came before any answer was read. It covers \frN{} new places, with the recorders' own clocks, a 300~s slack and a 20~km square. The best of three locators places clear-pass clips within 1~km at \frDpK{} places, where a uniform guess predicts \frExpU{} ($p=$~\frBestPc{} after correcting for the pick). With the wrong time or swapped audio it places none. The other two hit \frUnhK{} and \frRefK. On \aoBThreeN{} clips from \phThreePlN{} places of a GPS-timed network, two locators place \aoBThreeRiskPct\% and \aoBThreeDpPct\% within 1~km. By place, a unit chosen after that run, they hit \phThreeRefPlK{} and \phThreeDpPlK{} of \phThreePlN{} where a uniform guess predicts \phThreePlExp. On \eeN{} random public recordings, picked and cut without the truth in a time-bounded run, the three locators answer at most \eeAnsMax{} times and place at most \eeHitMax. We release AircraftWitness, a benchmark, and pass labels for \scRecM{} recordings.
\end{abstract}

%% file: sec/intro.tex
\section{Introduction}
\label{sec:intro}
Archives, field science and evidence rely on the place and time a recording states. Recorder clocks without GPS can drift by seconds a day~\citep{oad2022manual}, and stated coordinates in occurrence archives can be wrong~\citep{zizka2019}. Published outdoor audio can also reveal more about its place than its publisher intended. Aircraft overhead are one way this happens.

\iffig{doppler}{\begin{wrapfigure}{R}{0.4\linewidth}\centering\vspace{-\intextsep}
\includegraphics[width=\linewidth]{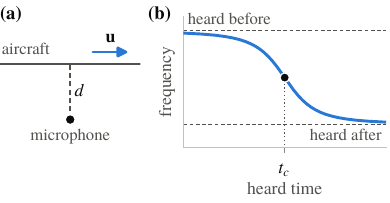}
\caption{(a) A pass drawn in its plane through the microphone, with velocity $\mathbf u$ and closest distance $d$. (b) The heard tone falls fastest at $t_c$, when the aircraft is nearest.}\label{fig:doppler}
\end{wrapfigure}}{}
Propeller aircraft and helicopters radiate tones whose pitch falls as they pass (\cref{fig:doppler}). Community networks archive the ADS-B positions that equipped aircraft broadcast~\citep{adsblol}. For a candidate place and time, the archive predicts the pass curves a microphone there would hear, and the best-matching place is the answer. Public demonstrations and open tools show that this works on cases their authors chose (\cref{sec:related}). How often it works on recordings that no one chose has not been measured.

We measure it in a registered test, timestamped before any audio and amended only before any answer was read, at \frN{} places new to every locator, with the recorders' own clocks and a 300~s slack. The best of three locators places clear-pass clips within 1~km at \frDpK{} of the \frN{} places, where a uniform guess predicts \frExpU, and none with the wrong time or swapped audio (\cref{sec:res-fresh}). An earlier blind test on \aoBThreeN{} recordings from \phThreePlN{} places in one GPS-timed network gives \aoBThreeRiskPct{} and \aoBThreeDpPct{} of every 100 within 1~km for two locators (\cref{fig:central}). In a time-bounded run on random public recordings, almost none are placed. The contributions are:
\begin{enumerate}[nosep,leftmargin=*]
\item The first measurement, to our knowledge, of how often aircraft place a recording, registered and timestamped before any audio, amended only before any answer was read, and scored by place against chance and controls (\cref{sec:res-fresh}).
\item The measured rates on a GPS-timed network and on random public recordings, xeno-canto, live streams and a park archive (\cref{sec:results}).
\item \dsName, a benchmark with the blind test's squares and an evaluator, and ADS-B pass labels for \scRecM{} recordings (\cref{sec:bench}).
\end{enumerate}

%% file: sec/related.tex
\section{Related Work}
\label{sec:related}
\paragraph{Locating recordings by their aircraft.} The approach is established. A public demonstration reports locating a security-camera recording from two aircraft passes~\citep{colsto2026doppler}. The open tools dopplerprint~\citep{dopplerprint2026} and DOPPLERGEIST~\citep{dopplergeist2026} search a region and start time for the place whose predicted pass curves fit a recording. Unheard~\citep{unheard2026} disclaims this search, yet its pinned commit contains one. AeroDoppler~\citep{aerodoppler2026} reports answering 18 of 29 Livermore phone recordings with a median error of 391~m. It gets the time within 2~s and a 4~km square, and we do not run it. These demonstrations use cases their authors chose or whose answers they knew, and some include controls. They show that the approach can work, but none measures how often it works over many recorders with held-out truth.

\paragraph{Aircraft passes in audio.} Reading a pass from the Doppler curve of its propeller tone is classical~\citep{ferguson1992doppler}. An earlier dataset matches the Doppler shifts of 40 smartphone-recorded passes to their flight tracks for speed and height~\citep{pathak2026livermore}. AeroSonicDB pairs ADS-B-triggered aircraft recordings with hand-checked audible events at three sites~\citep{downward2023aerosonicdb}.

%% file: sec/method.tex
\section{Measuring the Rate}
\label{sec:method}\label{sec:dataset}\label{sec:bench}
\paragraph{The witness.}\label{sec:m-doppler} An aircraft at $\mathbf p(\tau)$ with velocity $\mathbf u(\tau)$ emits a tone $f_0$, and the microphone is at $\mathbf r$. With range $R(\tau)=\lVert \mathbf p(\tau)-\mathbf r\rVert$ and sound speed $c$, sound emitted at time $\tau$ arrives at
\begin{equation}
t(\tau)=\tau+R(\tau)/c
\label{eq:arrival}
\end{equation}
with frequency
\begin{equation}
f\bigl(t(\tau)\bigr)=\frac{f_0}{1+\dot R(\tau)/c},\qquad \dot R=\frac{\mathbf u\cdot(\mathbf p-\mathbf r)}{R}.
\label{eq:doppler}
\end{equation}
For a straight pass, the curve gives the speed, the tone, and the time and distance of closest approach~\citep{ferguson1992doppler}.

\paragraph{Packets and hits.} Each test packet holds five minutes of daytime audio in which the tonal ridge detector finds a clear falling pass (\cref{sec:a-tests}). It gives the published start within 60~s, a 20~km square around the truth with a random offset, and the archived flights for that square and hour. An answer within 1~km of the published position is a hit, and a missing answer is a miss. At its pinned commit Unheard searches from the date and a 100~km square, and in the rerun it gets the same square and start time as the others.

\paragraph{Chance and controls.} Chance is the expected number of hits from a uniform guess in each recording's square. It counts the whole 1~km disk, so it is an upper bound. On the Australian Acoustic Observatory (A2O), recorders at one place share one sky. There a moment is one place at one start time, and a place holds up to four recorders within 5~km. Moments test whether a method beats chance, and places test how widely it works. Elsewhere the unit is the recorder (\cref{sec:a-protocol}). Each $p$-value tests how many units have a hit against chance. The square's centre is a baseline without audio. Controls rerun the same packets with the clock moved 5~min or 2~h and with another site's audio.

\paragraph{Blind runs.} A dated commit records each planned test and fixes its code by hash before it runs. No outside registry is used, so we call a test pre-registered only when an external timestamp predates its first run. Most tests were amended after their first run (\cref{tab:protocol}). The locator receives packets under random identifiers, and a separate automated verifier holds the truth and publishes only aggregates. No independent party ran or scored any test. The recorders' positions are public, so blindness rests on the fixed code and the verifier.

\paragraph{Sets.} The test set is A2O's batch 3, with \aoBThreeN{} packets from \aoBThreeSites{} recorders at \phThreePlN{} places new to this work~\citep{roe2021a2o}. Development used A2O's first two batches, with \aoLocN{} and \aoLocBTwoN{} packets at \aoLocSites{} and \aoLocBTwoSites{} recorders, and \eclN{} clips from \eclSites{} Eclipse Soundscapes recorders~\citep{eclipsesoundscapes}. Wider tests search \benchYrN{} later xeno-canto recordings, whose starts are known to minutes, in a 20~km square. They also search \liveCzN{} live-stream windows and \zenN{} park recordings over a \liveLbBoxKm~km square. Last, a stratified random sample draws \prevN{} public recordings with a labelled pass from an earlier inventory (\cref{sec:a-corpus}).

\paragraph{Entries.} The open dopplerprint and Unheard are the benchmark's baselines, each at a pinned commit. Our reference locator (\cref{sec:a-search}) uses their standard pipeline with dopplerprint's open pass detector, and we release it with the benchmark. dopplerprint searches the circle around the square (\cref{sec:a-protocol}).

\paragraph{The benchmark.} \dsName{} holds \dsRecordings{} recordings from \dsSources{} sources, with places and times as each source publishes them. It also holds \dsPasses{} ADS-B proximity labels within \dsRadiusKm~km in \dsRecPass{} recordings (\cref{fig:dataset}). A label means an aircraft passed within range, not that it was heard (\cref{sec:a-detect}). A wider run labels about \scRecM{} listed recordings, of which \scRecPassM{} have a pass (\cref{tab:corpus}). An evaluator scores answer files on the released test squares.

%% file: sec/results.tex
\section{Results}
\label{sec:results}\label{sec:exp}
\input{build/tab_main.tex}
\Cref{tab:main} gives the test set from the one-machine rerun and \cref{tab:mainmore} further rows. In both, a blank count or chance repeats the value above, and a blank recorder count marks a set not counted by recorder.

\subsection{The Blind Test}
\label{sec:res-field}\label{sec:res-bench}
\iffig{central}{\begin{wrapfigure}{R}{0.48\linewidth}\centering\vspace{-\intextsep}
\includegraphics[width=\linewidth]{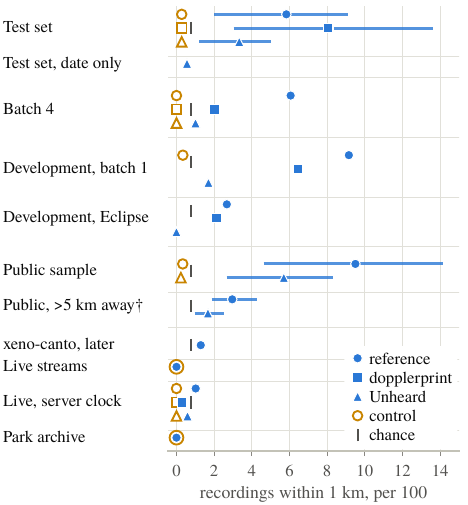}
\caption{Recordings placed within 1~km per 100 tested, with ticks for chance and open points for the higher wrong-time or wrong-audio control. Lines are 95\% intervals from resampling places. $\dagger$ post hoc.}\label{fig:central}
\end{wrapfigure}}{}
On the test set the reference places \aoBThreeK{} of \aoBThreeN{} packets within 1~km and dopplerprint places \aoBThreeDpK{}, with \aoBThreeDpBoth{} hits in common. The reference run was blind and pre-registered, and dopplerprint ran on the same packets afterwards. The registered test counted the \aoBThreeSites{} recorders as independent (\cref{sec:a-protocol}). We report moments and places instead, both chosen after the run. The two locators hit \phThreeRefMoK{} and \phThreeDpMoK{} of \phThreeMoN{} moments where chance predicts \phThreeMoExp{} ($p=$~\phThreeRefMoP{} and \phThreeDpMoP). They hit \phThreeRefPlK{} and \phThreeDpPlK{} of \phThreePlN{} places where chance predicts \phThreePlExp{} ($p=$~\phThreeRefPlP{} and \phThreeDpPlP). A moved clock or another site's audio gives the reference at most \aoBThreeCtlMaxK{} hits, near chance (\cref{tab:controls}).

A planned rerun on one machine gives every entry identical inputs and limits. The reference and dopplerprint keep the same hits. Unheard places \rrUnhK, at \phThreeUnhMoK{} moments ($p=$~\phThreeUnhMoP) and \phThreeUnhPlK{} places ($p=$~\phThreeUnhPlP), and at most \rrUnhCtlMax{} under either control. A separate blind and pre-registered test gives Unheard only the date and a 100~km square. It places \unhBThreeK{} of \unhBThreeN{} test packets and \unhAoK{} of \unhAoN{} in batch 1, beyond chance when recorders count as independent ($p=$~\unhBThreeP{} and \unhAoP). Batch 4 has \bFourN{} packets but is no test at new places, since \phFourPlUsed{} of its \phFourPlN{} places were used before (\cref{tab:main}, \cref{sec:a-tests}).

\ifhave{frN}{\subsection{The Registered Test}\label{sec:res-fresh} Its plan was timestamped before any audio, and its two amendments came before any answer was read. \ifhave{frOK}{A registered replication}{A registered new-place test} covers \frN{} new Eclipse places over 5~km from any place used before, each with up to six clear five-minute clips, one a day.\ifnum\frN<20{} Only \frN{} places passed screening by the stop, below the plan's minimum of 20, so this is a feasibility shortfall and not a replication, and its $p$ values are descriptive.\fi{} The start comes from the recorder's own clock, so every locator gets a 300~s slack. The place is the unit, and chance is the blind test's shift null. Of the \frN{} places, the reference hits \frRefK{} against \frRefExp{} by chance ($p=$~\frRefP), dopplerprint \frDpK{} against \frDpExp{} ($p=$~\frDpP) and Unheard \frUnhK{} against \frUnhExp{} ($p=$~\frUnhP). A uniform guess would predict \frExpU{} places, with the same verdicts for the open tools. Picking the best of the three, a threefold correction gives $p=$~\frBestPc{} against that guess. The clear-pass detector was used in developing the reference, so the reference is descriptive\ifhave{frOK}{ and each open tool is its own registered question at $p<0.05$. \ifdim\frDpPv pt<0.05pt\ifdim\frUnhPv pt<0.05pt Both replicate.\else Of the two, only dopplerprint replicates.\fi\else\ifdim\frUnhPv pt<0.05pt Of the two, only Unheard replicates.\else Neither replicates.\fi\fi}{.}  With the same slack, the post hoc rerun of the blind test hits \swRefSlLongPlK{} and \swDpSlLongPlK{} of its \phThreePlN{} places (\cref{sec:a-sweep}). Of the \frClips{} clips, the reference answers \frRefAns, dopplerprint \frDpAns{} and Unheard \frUnhAns, and \frDpHits{} of dopplerprint's answers are within 1~km. With the wrong time or another place's audio, dopplerprint hits no place, and \ifnum\frCtlMaxK=0 no entry hits one.\else no entry hits more than \frCtlMaxK{} of the \frN{} places.\fi

}{}\subsection{How Often It Works}
\label{sec:res-rare}
Per recording, the two locators place \aoBThreeRiskPct\% and \aoBThreeDpPct\% of test packets within 1~km. Resampling whole recorders gives \anYieldRefCi{} and \anYieldDpCi{}, and resampling the \phThreePlN{} places gives the coarser \bThreePlCiRef{} and \bThreePlCiDp{}. These rates hold for daytime five-minute clips with a pass the detector finds clear, at \anPopSensors{} recorders with flight coverage and \anPopPerLo{} to \anPopPerHi{} clips each. They are not rates for outdoor recordings in general.

\ifhave{eeN}{An earlier planned sample of \prevN{} public recordings with a labelled pass is a best case, since longer recordings were cut around a pass found with the true place (\cref{sec:a-corpus}). There the reference places \prevRefPct\% within 1~km, Unheard \prevUnhPct\% and dopplerprint, which covers only the contiguous US, \prevDpPct\%, against \prevExpPct\% by chance. Most hits come from places used in this work. After the run, more than 5~km from them, the reference places \prevOosRefPct\% of \prevOosN{} recordings, and \prevOosEcoRefK{} of its \prevOosRefK{} hits are on Ecosounds, whose operator also runs A2O's portal~\citep{ecosounds,a2oportal}, leaving \prevOosRestRefK{} of \prevOosRestN{} elsewhere.}{The planned random sample holds \prevN{} public recordings with a labelled pass (\cref{sec:a-corpus}). Recordings up to five minutes are used whole, and longer ones are cut around a pass found with the true place, so this is a best case. Under a later amendment the reference runs blind and places \prevRefPct\% within 1~km, with \prevRefCiPl{} from resampling places. Unheard places \prevUnhPct\% (\prevUnhCiPl) and dopplerprint, which covers only the contiguous US, places \prevDpPct\%, all against \prevExpPct\% by chance. With the wrong time or another recording's audio the reference places \prevWtPct\% and \prevWaPct\%. Most of these hits come from places used in this work, so after the run we kept the \prevOosN{} recordings more than 5~km from our A2O, Eclipse, park, live-stream and Livermore test places. There the reference places \prevOosRefPct\% (\prevOosRefCi) and Unheard \prevOosUnhPct\% (\prevOosUnhCi). Of the reference's \prevOosRefK{} hits, \prevOosEcoRefK{} are on Ecosounds, whose operator also runs A2O's portal~\citep{ecosounds,a2oportal}. The other archives give it \prevOosRestRefK{} of \prevOosRestN{} (\prevOosRestRefPct\%). As an illustration, \prevSnapPassPct\% of an earlier inventory has a labelled pass, so the \prevOosRefK{} hits away from our sites among all \prevN{} come to \prevFunOos{} per 100 inventory recordings.}

\ifhave{eeN}{\paragraph{Random public recordings.}\label{sec:res-e2e} A registered test draws \eeN{} random public recordings with a published place and start inside the flight archive's dates. No pass label or true place picks a recording or its clip. Recordings up to five minutes are used whole, and longer ones are cut where the detector finds a pass, or at random. Each locator gets the stated start within 60~s and a 20~km square. A failed fetch, crash, empty answer or run unfinished at the processing stop is a miss. Three amendments, before any score, changed how 343, 121 and 258 recordings were downloaded, and the last also moved the stop later. Of the \eeN{}, \eeReady{} gave audio and up to \eeUnrun{} were unfinished at the stop. The reference answers \eeRefAns{} times and places \eeRefK{} within 1~km\ifhave{eeRefCi}{ (95\% interval over days \eeRefCi)}{}. Unheard answers \eeUnhAns{} times and places \eeUnhK. dopplerprint answers \eeDpAns{} times and places \eeDpK. With the wrong time or another recording's audio, which finished for fewer recordings by the stop, no entry places more than \eeCtlMaxK. A guess on every recording would place a nominal \eeExpPct\%, and the plan sets no test. So in this time-bounded run, the locators almost never answered on random public audio.

}{}\subsection{Archives and Streams}
\label{sec:res-nil}\label{sec:res-live}
The reference places \eclK{} of \eclN{} Eclipse clips, at \eclSitesHit{} of \eclSites{} sites ($p=$~\eclSitesP). Post hoc, pooling by recorder-day makes the sites stand out ($p=$~\eclPoolSitesP, \cref{sec:a-tests}). On later xeno-canto recordings it places \benchYrK{} of \benchYrN{}, about chance. US phone recordings on iNaturalist give \glInusKm{} of \glInusN{} within 1~km against \glInusExpAll{} by chance (\cref{sec:a-global}). That test was neither planned nor blind, and its flight cut centres on the truth.

A test planned before the windows were drawn uses \lvTwoN{} live-stream windows from \lvTwoStreams{} open microphones, given a 20~km square and the server's start time. The reference places \lvTwoRefK{} within 1~km, dopplerprint \lvTwoDpK{} and Unheard \lvTwoUnhK{}. None significantly exceeds the \lvTwoExp{} hits from uniform guessing, and none hits under either control. The registered test counts streams hit against the shift null of batch 3. It passes the reference ($p=$~\lvTwoRefStP) and dopplerprint ($p=$~\lvTwoDpStP), and Unheard is borderline ($p=$~\lvTwoUnhStP). The registered set had \lvWcSt{} webcams, which we left out after scoring because their publisher grants no permission to record. That departs from the plan, and with them the test gives $p=$~\lvRegRefStP, \lvRegDpStP{} and \lvRegUnhStP{} (\cref{sec:a-live}).

\subsection{What an Answer Is Worth}
\label{sec:res-worth}
An answer alone does not show whether it is right. On the test set about one in \aoBThreeAnsPerHit{} answers from either locator is a hit (\cref{tab:main}). Answers an entry marks sure hit more often across its tests, but about half still miss by over 1~km, in a comparison made after the run (\cref{fig:worth}).

\subsection{Reading Recorder Clocks}
With the place known, a line fitted through the passes gives a recorder's clock offset and drift. On the Livermore phone recordings, single passes read the phones' own clock with a robust standard deviation of \clockPassSd~s. No GPS clock confirms these readings. On \clockLongSites{} GPS-timed A2O recorders, whose public positions are rounded, the passes resolve \clockLongRes{} of the \clockLongFit{} clocks with enough passes (\cref{sec:a-clock}).

%% file: build/tab_main.tex
\begin{table}[t]\centering\small
\caption{Recordings placed within 1 km. Chance is the hits uniform guesses in each square give. Recorders counts those with a hit. $\dagger$ post hoc.}\label{tab:main}
\setlength{\tabcolsep}{5pt}\renewcommand{\arraystretch}{0.92}\begin{tabular}{@{}llrrrrr@{}}\toprule
Set & Entry & Recordings & Answered & Hits & Chance & Recorders \\
\midrule
Test set & reference & 360 & 98 & 21 & 2.83 & 13\\
 & dopplerprint\rlap{\smash{$^\dagger$}} &  & 157 & 29 &  & 13\\
 & Unheard &  & 92 & 12 &  & 9\\
 & square's centre\rlap{\smash{$^\dagger$}} &  & 360 & 1 &  & 1\\
\addlinespace[1.5pt]
Test set, date only & Unheard & 360 & 309 & 2 & 0.11 & 2\\
\addlinespace[1.5pt]
Batch 4 & reference & 99 & 16 & 6 & 0.78 & 4\\
 & dopplerprint &  & 28 & 2 &  & 2\\
 & Unheard &  & 4 & 1 &  & 1\\
\addlinespace[1.5pt]
Development, batch 1 & reference & 295 & 184 & 27 & 2.32 & 4\\
 & dopplerprint\rlap{\smash{$^\dagger$}} &  & 200 & 19 &  & 5\\
\addlinespace[1.5pt]
Development, Eclipse & reference & 187 & 48 & 5 & 1.47 & 3\\
 & dopplerprint\rlap{\smash{$^\dagger$}} &  & 83 & 4 &  & 4\\
\addlinespace[1.5pt]
xeno-canto, later & reference & 155 & 58 & 2 & 1.22 & \\
\addlinespace[1.5pt]
Live streams, server clock & reference & 686 & 68 & 7 & 5.39 & \\
 & dopplerprint &  & 47 & 2 &  & \\
 & Unheard &  & 54 & 4 &  & \\
\addlinespace[1.5pt]
Park archive & reference & 232 & 101 & 0 & 0.01 & 0\\
\bottomrule\end{tabular}
\end{table}

%% file: sec/meaning.tex
\section{What the Rate Means}
\label{sec:recordists}\label{sec:res-defense}
\paragraph{For recordists.} Most hits in these tests fall on fixed recorders near traffic whose time and rough area are published (\cref{sec:res-field}). The rate is per recording, so a recorder that publishes many recordings is more likely to get at least one right answer among many wrong ones. On the test set \aoBThreeSitesHit{} of \aoBThreeSites{} recorders have at least one hit, from \anPopPerLo{} to \anPopPerHi{} recordings each (\cref{sec:res-rare}). Most noise suppressors we tested keep most of the passes the detector finds, but after RNNoise and a tone remover it finds almost none (\cref{sec:a-defense}).

\paragraph{For archives.} Run forward, the same witness checks a recording's stated place and time. On US recordings in a bird-sound archive it favours the stated values over shifted ones for about 1 in \auditConfOneIn{} recordings in which its detector finds a pass. It reaches a decision on only \auditFalseDecPct\% of held-out decoys moved 150~km (\cref{sec:res-tag}).

\paragraph{Why the rate is small.} A hit needs an answer. The reference finds no matching flight for \aoBThreeNoMatch{} of the \aoBThreeN{} test packets and no pass in \aoBThreeNoPass{}. At every reference hit the closest archived aircraft is a propeller aircraft or helicopter within 10~km (\cref{sec:a-hit}).

%% file: sec/ethics.tex
\section{Limitations and Ethics}
\label{sec:ethics}\label{sec:limitations}
The blind test covers \phThreePlN{} places in one GPS-timed network. By moment it is far above chance. By place it is \phThreeRefPlK{} and \phThreeDpPlK{} of \phThreePlN{} against \phThreePlExp, which alone does not settle it, and both units were chosen after the run. \ifhave{frN}{The new-place test assumes each device clock is within 300~s from a fleet model, not measured per recorder. }{}\ifhave{eeN}{}{The public-recording rate is a best case. Every sampled recording has a pass found with the true place, longer ones were cut around it, and most hits come from one archive. }\ifhave{eeN}{\ifhave{frOK}{}{A test at 20 or more new places comes next. }}{\ifhave{frOK}{A rerun that picks clips by detector alone comes next. }{A rerun that picks clips by detector alone and a test at 20 or more new places come next. }}ADS-B misses aircraft without it, which US rules require only in some airspace~\citep{faa91225}. The model assumes straight-line sound at one speed. The labels are ADS-B proximity, not hand labels\ifhave{audN}{, and audibility rules disagree}{}.

Anyone can already run this search with open tools~\citep{dopplerprint2026,dopplergeist2026}. We release the benchmark code, the dataset and our reference locator. Searches use places that are already public. The exception is the Livermore clips (\cref{sec:a-clock}). Their publisher's distances imply the place. The release holds the code and inputs that recover it, but we print no address. Results use public recordings under their terms (\cref{sec:a-licences}), except webcam audio, which is recorded without permission and retained only in the registered stream test (\cref{sec:a-live}). There is no contact with people, and other personal data are limited to what the sources publish.

%% file: sec/conclusion.tex
\section{Conclusion}
\label{sec:conclusion}
Aircraft can place a recording, and this paper measures how often. In a test registered before any audio and amended only before any answer was read, at \frN{} new places with the recorders' own clocks, the best of three locators places clear-pass clips at \frDpK{} places where a uniform guess predicts \frExpU, and none with the wrong time or audio. The other two hit \frUnhK{} and \frRefK. On a GPS-timed network with a tight time prior, two locators place \aoBThreeRiskPct\% and \aoBThreeDpPct\% of clear-pass clips within 1~km. On \eeN{} random public recordings, picked and cut without the truth, the three locators answer at most \eeAnsMax{} times and place at most \eeHitMax. These rates measure today's tools on these recordings, and a better locator could find more. AircraftWitness makes this measurable for anyone, with pass labels for \scRecM{} recordings and a benchmark to score new locators.

%% file: sec/statements.tex
\section*{Release and Acknowledgements}
\label{sec:statements}
Our code, reference locator, dataset, registered plans and per-recording results are at \relUrl. Everything in the paper is under CC0~1.0 unless stated otherwise. Third-party data and code keep their own licences (\cref{sec:a-licences}). This work was supported in part through the NYU IT High Performance Computing resources, services, and staff expertise. Eclipse Soundscapes is supported by NASA award No.~80NSSC21M0008. AI tools assisted with this work, including the literature review and cleaning the code and dataset of private identifiers (\cref{sec:ethics}). The author reviewed and verified all content and takes responsibility for it.

%% file: sec/appendix.tex
\section{Reference Implementation}
\label{sec:a-search}\label{sec:m-search}\label{sec:m-audit}
The entry uses the open tools' standard method (\cref{sec:related}). The open dopplerprint detector finds passes as ridges in a narrow-band spectrogram~\citep{dopplerprint2026}. Each gives $t_c$, tone, speed and distance with intervals. A track explains a pass if its closest distance and speed fall inside these intervals (\cref{eq:arrival,eq:doppler}). With the clock unknown, two passes' spacing must match within 12~s. With the start time given, \aoBThreeOnePassK{} of the test set's \aoBThreeK{} hits use a single pass. Candidates that predict a loud pass the recording lacks are dropped. The rest rank by the rms frequency distance between the predicted curves and the ridges traced around them, over the fundamental and harmonics. Date-only runs instead rank by excess. It is evidence minus the sum of the mean and two standard deviations of the same place's evidence with the clock moved. The search spans $f_0$ within $\pm$12\% and one clock offset within $\pm$4~s in 0.25~s steps. The ten best flight pairs are re-searched on a 50~m grid. In the unknown-clock search, a candidate is dropped if another aircraft passes within 4~km between the two passes, more than 40~s from either, and the recording shows no such pass. Tracks are interpolated over gaps of at most 60~s. Airborne aircraft with at least four positions are kept, including unknown types. Date-only seed passes below \dateFJet~Hz never match jets. Heights are geometric, or barometric and corrected by nearby aircraft reporting both. Grid points sit 2~m above GLO-30 terrain, or GLO-90 on A2O. One geoid undulation applies across the region. Sound speed follows from an assumed air temperature. No grid node falls on the truth. After the search, the truth is used to report rank and error. From the date alone, the area is the bird prior's region (\cref{sec:m-birds}).
\paragraph{Check.} The forward check takes a stated place $\mathbf r_0$ and start $T_0$. It predicts every pass the archive implies there and scores the spectrogram along the predicted curves. It tests the stated values against likely slips. These are whole-hour shifts from $-13$ to $+13$~h, a day either way, and nearby places. Each slip leaves the clock free within 5~min. A tone group matches an archived pass with speed within \auditRuleV\% of the fitted one, closest distance \auditRuleDlo{} to \auditRuleDhi{} times the fitted one and a curve fit within \auditRuleZ{} noise units. A recording is confirmed when a match at the stated values is also the group's best fit under every slip. It is shifted when only a slip matches and two groups agree on it within \auditRuleW~s or its fit beats the group's others \auditRuleGap-fold. Confirmation means the stated values beat these slips, not that they are exact.

\section{Detecting Passes}
\label{sec:a-detect}
The dopplerprint detector finds \dsDetNearK{} of the \dsDetNearN{} proximity labels within \dsDetNearHi~km and \dsDetFarK{} of \dsDetFarN{} at \dsDetFarLo{} to \dsDetFarHi~km, over \dsDetRec{} recordings. Post hoc, over \anDetEqPairs{} label and decoy pairs in \anDetEqRec{} recordings, it matches \anDetEqRecall\% of labels and \anDetEqDecoy\% of decoys moved a minute or two. Both the label and its decoy must lie inside the audio. A one-sided sign test by recording gives $p=$~\anDetEqP. Unmatched detections total \dsDetUnmatchedHr{} an hour.

Of \audN{} passes with public audio sampled by stratum, \audDet{} can be scored. A pass is undetermined if its closest approach lies outside the audio, the audio is bad or missing, or too few quiet windows surround it. Each rule compares the 20~s around closest approach with nearby quiet windows. A frozen band-energy rule flags \audPct\%, weighted to the corpus, with a 95\% interval of \audCi. It flags \audNearPct\% within 2~km, \audMidPct\% at 2 to 5~km and \audFarPct\% at 5 to 10~km. The PANNs CNN14 classifier~\citep{kong2020panns} flags \audClfPct\%. Quiet windows are scored by the same rule. On them the two flag \audDecoyPct\% and \audClfDecoyPct\%. This is a background rate, not a checked false-flag rate. Unweighted, the two agree on \audAgreePct\%, with $\kappa=$~\audKappa. The weighted share rests on few passes in the heaviest strata. It assumes undetermined passes resemble determined ones within each stratum. Unweighted, it is \audUnwPct\%.

\section{Prospective Blind Test}
\label{sec:a-live}
The reference is frozen and its hash logged with a separate verifier before any audio capture. The scoring rule changes after capture begins, and each version is hashed. About an hour in, one version seals the planned arm of \liveSamplePct\% of later ten-minute slots by keyed hash. Later versions do not change it. Windows are from \liveScStreams{} Locus Sonus open microphones~\citep{locustream}. Each packet has a random identifier, audio, a start with $\pm$\liveSlackS~s slack and a square placed at random around the truth. Flights come from the live APIs of adsb.fi~\citep{adsbfi} and adsb.lol, polled over areas that reveal no stream. The locating worker never reads the stream list or the truth. Blind controls give the same audio with flights from two hours before, or another stream's region and flights. A stream's squares together narrow its place, so blindness requires the frozen locator to read one packet at a time. The rule, windows and truth were timestamped publicly before any score was shared. The verifier scored them privately first. The places were already public. In the \liveLbBoxKm~km square all \liveWindows{} windows give \liveK{} hit against \liveExp{} by chance. In the design's own \liveLtBoxKm~km square the planned arm gives \liveLtCzK{} hit in \liveLtCzN{} windows and all \liveLtN{} windows give \liveLtK. Every control gives none.
\paragraph{Server-clocked arm.} A second arm gives each window a 20~km square and the start time from the stream's server, including the server's delay. Against a clock checked with time servers, the reference's start is off by a median \lvTwoClkRef~s, dopplerprint's by \lvTwoClkDp~s and the server's by \lvTwoClkNull~s. As registered, streams placed coarser than 300~m are left out. Packets without flight rows are misses for every entry. Some streams' places were in the first arm's released dataset, which no entry reads. The registered set of \lvRegN{} windows from \lvRegStreams{} streams also held \lvWcWin{} windows from \lvWcSt{} webcams of a commercial service. None hit. These and their \lvWcCtl{} control windows were excluded after scoring, since their publisher grants no permission to record. The rest are open microphones whose owners publish their places: \lvTwoLsSt{} of Locus Sonus, \lvTwoApSt{} of radio aporee~\citep{aporee}, \lvTwoCfSt{} of Cyberforest~\citep{cyberforest} and \lvTwoIcSt{} other Icecast streams. With the wrong time (\lvTwoWtN{} windows) or another stream's audio (\lvTwoWaN), no entry places any.

The registered test counts streams hit against batch 3's shift null. On the open microphones the reference hits \lvTwoRefSt{} streams ($p=$~\lvTwoRefStP), dopplerprint \lvTwoDpSt{} ($p=$~\lvTwoDpStP) and Unheard \lvTwoUnhSt{} ($p=$~\lvTwoUnhStP). On the registered set the tests give $p=$~\lvRegRefStP, \lvRegDpStP{} and \lvRegUnhStP. Corrected post hoc for three entries, they are \lvRegRefStPBonf, \lvRegDpStPBonf{} and \lvRegUnhStPBonf. Under a uniform guess on every window, no entry beats chance ($p=$~\lvTwoRefUniP, \lvTwoDpUniP{} and \lvTwoUnhUniP). Post hoc with streams as units, uniform chance gives $p=$~\lvTwoRefStUniP, \lvTwoDpStUniP{} and \lvTwoUnhStUniP. Of the reference's \lvTwoRefK{} hits, \lvTwoRefTopK{} fall on one stream. A capture stall leaves \lvRegStallN{} registered windows in one of the \lvRegBands{} half hours. Rows rerun after a logged fault, before scoring, change no count.

\section{Recorder Clocks}
\label{sec:res-clock}\label{sec:a-clock}
A test planned in a dated commit uses up to four days at each of \clockLongSites{} of A2O's GPS-timed recorders. The passes resolve \clockLongRes{} of the \clockLongFit{} with enough passes. They put their clocks \clockLongLo{} to \clockLongHi~s from labels naming unverified nominal starts. \clockLongBlocksRes{} of \clockLongBlocks{} two-hour files hold a clock line with twice the passes of any other. No site's pooled fit does. Offsets of both signs reject one common lag ($\chi^2=\clockLongChi$ on \clockLongDof{} degrees of freedom). Yet over the eclipse recorder's passes, 200~m of position error moves an offset by up to \clockBudAoOff~s, in opposite directions for opposite headings. A label moved 3~s moves the fit \clockLongCtl~s at \clockLongCtlN{} sites. This control's expected sign was corrected after its first results were read. Opposite traffic directions give estimates that agree on \clockLongHeadAgree{} of \clockLongHeadDec{} site-days. These site-days test the rounded positions. This comparison is post hoc. On \aoSites{} other GPS-synchronised recorders at their rounded public positions, the passes resolve \aoResolvedN{} of the clocks. Pass-read clocks are thus checked only against phones' clocks and file gaps. No test checks whether an exact position and two-way traffic suffice.

With the place fixed, each pass $e$, matched to track $j_e$, gives an offset $\hat t_{c,e}-t_c^{j_e}$ between its heard and predicted closest-approach times. A recorder whose oscillator runs fast by $\epsilon$ reads $\hat t_{c,e}=(1+\epsilon)\,t_c^{j_e}+\Delta$. A RANSAC line through the passes gives offset $\Delta$ and drift $\epsilon$. Per-pass errors correlate within a day, so fit errors are optimistic. Nothing bounds clock changes between passes. An archive lag of $\sim$\clockLag~s is not subtracted. It is assumed from the median Livermore offset against the phones' whole-second labels. The $\pm$ values are fit standard errors only, and the bands are fit-only 95\% intervals.

\optfig[!htbp]{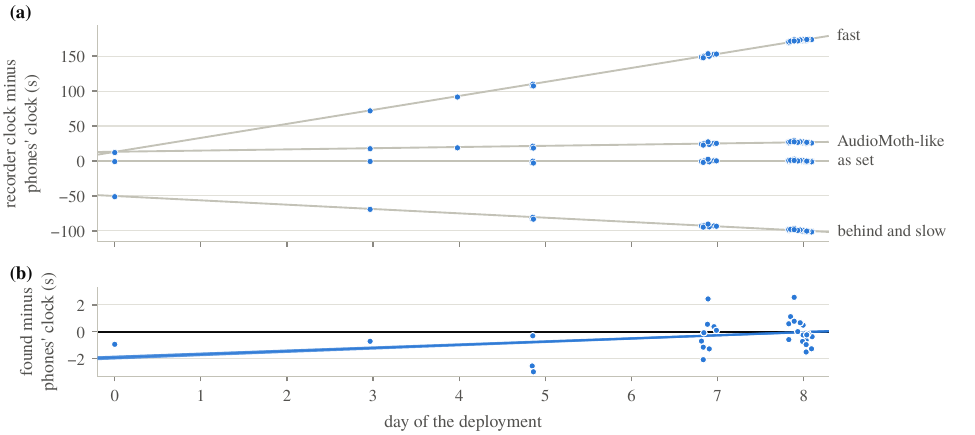}{Livermore passes against the phones' clock, which is not independent UTC. Grey shows injected clocks, and blue the four fits minus their injected clocks.}{fig:clocklv}
The \clockClips{} Livermore clips with an aligned start form one controlled deployment (\cref{fig:clocklv}). Its panel (b) shows found minus the phones' clock, clipped at $\pm$3.3~s. Their starts come from the phones' clock in whole seconds. Its publisher gives no recorder position~\citep{pathak2026livermore}. A fit matches each ADS-B closest-approach distance to the dataset's per-pass value, and the code and inputs for the fit are in the release. \baseLivN{} passes agree to \baseLivRms~m rms. \baseLivOut{} fit a point \baseLivOutKm~km away and are scored against it. Three injected clocks shift each clip's start by $\Delta+\epsilon(t-t_\mathrm{set})$ and resample its audio by $\epsilon$. For these and the clips as set, the fit gives each clock within \clockOffsetErr~s rms and the drift within \clockDriftErr~s per day. This checks the fit's arithmetic, not the readings. As set, the fitted offset is $\clockLivNatOff\pm\clockLivNatSd$~s from the phones' clock.

\optfig[!htbp]{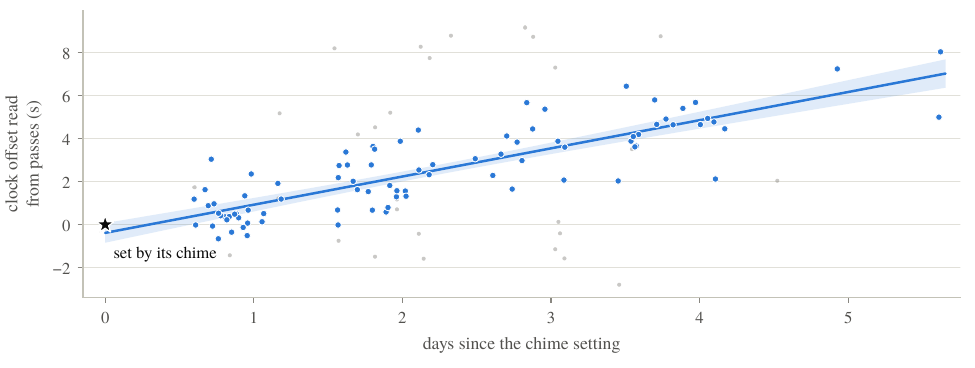}{The clock of ESID\#278 read from aircraft passes, with no GPS or independent truth. Passes on the line are blue, the band is the fit's 95\% interval and the star is the chime setting.}{fig:clock}
An Eclipse Soundscapes AudioMoth without GPS stays outdoors \clockEsDays{} days~\citep{eclipsesoundscapes,eclipse2026esid278}. Of its \clockEsGated{} gated passes, \clockEsInliers{} fall on one clock line (\cref{fig:clock}). The drift is \clockEsDrift{}~$\pm$~\clockEsDriftSd~s per day, within the maker's typical 2~s a day~\citep{oad2022manual}. Back at the phone chime that set the clock, the fit gives \clockEsAnchor{}~$\pm$~\clockEsAnchorSd~s, near the zero a chime-set clock reads. The chime's own error is not measured. Without aircraft, two-clock fits on \clockGapN{} of the project's recorders predict their file gaps ($r=$~\clockGapR). Right fits require this.

\section{The Archive Check}
\label{sec:res-tag}
The check's threshold is set with \auditCalibXcK{} xeno-canto confirmations in view, before any held-out decoy is scored. Resetting it without each recordist in turn leaves \anLoo{} of its \auditConfK{} confirmations. The original \auditConfK{} confirmations come from \auditXcConfRec{} recordists, and two give \auditXcConfTopTwo. \auditXcConfOneAc{} rest on one aircraft each. Only \auditXcConfMinK{} fit within a minute of the stated time. On Livermore it decides few shifted clocks but corrects each it decides. These decided shifted cases use \auditInjDecRec{} distinct recordings in sample. On xeno-canto it confirms stated times and places far more often than at decoys. It shifts no more often. Its shifts cannot be scored there (\cref{tab:audit}).
\input{build/tab_audit.tex}
In \cref{tab:audit}, a shifted recording counts once per shift. The indented row scales the xeno-canto decoys' rates to these recordings. The last row pools the held-out decoys from xeno-canto and US and European iNaturalist, each moved 150~km. The later European xeno-canto decoys (\cref{sec:a-global}) are not in it.

\section{Benchmark Protocol}
\label{sec:a-protocol}
\optfig[!htbp]{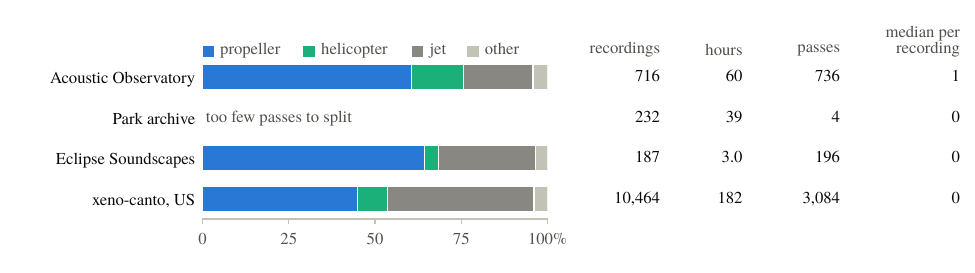}{Sources with released pass labels and their aircraft mix within 10~km. Counts and valid-cut medians appear at right.}{fig:dataset}
In \cref{fig:dataset}, medians use only recordings with a valid flight cut. Other includes unknown aircraft types. Livermore and the live streams have no released labels and make up the rest of the count in \cref{sec:method}. 
T1, the locate task, gives audio, start time and a square. The square is never centred on the truth. xeno-canto squares are random and stored before the runs. Livermore offsets are seeded by recording identifier and can in principle be inverted, so those known-place cases do not establish blindness. How an attacker obtains a square is not tested. Methods receive the same square and supplied flight cuts per task, subject to dopplerprint's adjacent-day exception in \cref{sec:a-tests}. Time priors are $\pm$5~min for xeno-canto and $\pm$2~s for Livermore. Unheard's date-only runs instead receive a 100~km square. dopplerprint takes a circle and receives the circle around the square. Runs have 30-minute limits, and dopplerprint on A2O an hour. Rows using given pass windows are labelled oracle and left out. The reference is fitted on Livermore and developed before any test. Only the batch 3 rerun gives every method common inputs. The dataset holds the test squares and an evaluator that scores answer files against them and the public positions. Scoring each square's centre reproduces the test set's centre row in \cref{tab:main}. Released truths are public, so new scores on them are development results. Only a sealed test on new recorders is blind.
\paragraph{Chance.} Unless stated otherwise, every number treats all recordings as uniform guesses. This is the most conservative chance model. The registered null keeps each answer's offset in its square and scores it against other recorders' truths. It gives the reference \aoBThreeShiftExp{} recorders ($p=$~\aoBThreeShiftP). The registered test counts the test set's \aoBThreeSites{} recorders as independent. Of these, \aoBThreeSitesHit{} hold a hit against \aoBThreeSitesExp{} by uniform chance ($p=$~\aoBThreeP), under the first registered rule and its amended form. The recorders sit at \phThreePlN{} places, so the text counts moments and places. Under the registered null, with the clock moved 5~min later and earlier, $p=$~\aoBThreeCtlP. Both early hits lie on the 60~s bound. Post hoc and untested, we read them as an edge effect. Dropping answers on that bound from every arm, the stated time still hits at \anEdgeStSites{} recorders and the moved clocks at \anEdgeSitesW.
\input{build/tab_controls.tex}
\input{build/tab_protocol.tex}
In \cref{tab:protocol}, registered is when the plan was committed. Timestamped is when the OpenTimestamps calendars received it. A test is pre-registered where that timestamp precedes its first run. Otherwise it is planned in a dated commit. Amendments count every change to the plan, and the last column those after the first run. The rerun, batch 4, the random sample and the server-clocked live arm are planned in dated commits.
\optfig[!htbp]{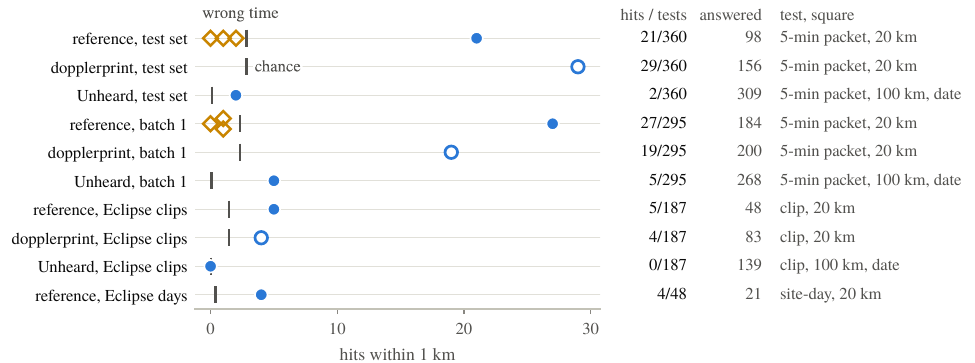}{First-run field tests show hits, chance and wrong-time controls. Filled dots mark planned tests, and open dots mark post hoc tests.}{fig:field}
\optfig[htbp]{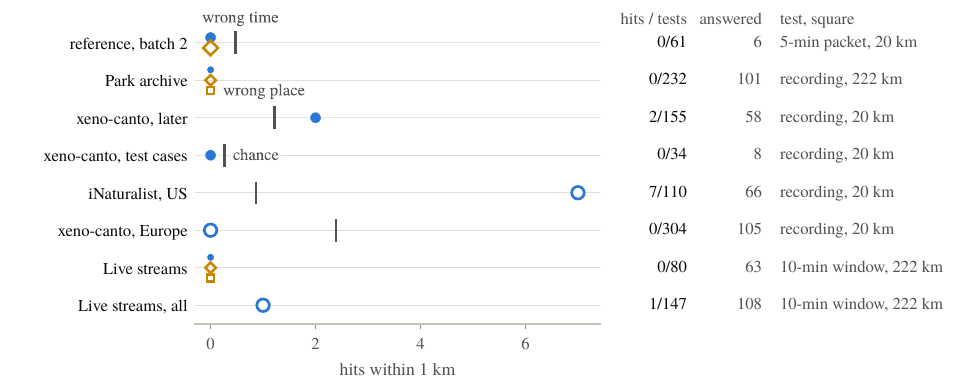}{Other field tests use the symbols of \cref{fig:field}, with squares for wrong places. Unreported controls are absent.}{fig:fieldmore}
In \cref{fig:field,fig:fieldmore}, ticks show chance and diamonds the wrong time. Planned does not necessarily mean pre-registered. The rerun in \cref{tab:main} has 157 dopplerprint answers against 156 in \cref{fig:field}, with the same hits. Controls use the same recordings or other live windows and may hold fewer recordings than their row.
\paragraph{Corpus.}\label{sec:a-corpus} \Cref{tab:corpus} counts the corpus. The random sample and the inventory rate in \cref{sec:res-rare} use a previous inventory without iNaturalist. Passes occur in \prevSnapPassPct\% of its recordings with public audio. Adding iNaturalist raises the share to \scPubPassPct\%. Sampled recordings up to 300~s are used whole. Longer ones are cut to 300~s, with a keyed offset placing the labelled pass's closest approach 30 to 270~s in. Each recording receives a 20~km square by batch 3's rule, its claimed start within 60~s and the square's flights. The wrong-time control shifts by 2~h. The wrong-audio control uses another sampled recording's audio from over 60~km away. Where a source gives no duration, a pass counts if its closest approach falls within 300~s of the start. The audio may be shorter. Two checks are not pre-registered. One place \prevTopDevKm~km from a development site gives \prevRefTopK{} of the reference's \prevRefK{} hits. Without it the rate is \prevRefNoTop\%. A2O recordings number \prevAoNs{} of the \prevN{} and give \prevRefAoK{} of its hits. Elsewhere it places \prevRefNoAoPct\%, and none of \prevRefXcN{} from xeno-canto. Post hoc, one place gives \prevTopK{} of Unheard's \prevUnhK{} hits. Without it, its rate is \prevUnhNoTop\%. Audio not public is available only on request or not at all. The labels ship without the recordings' places and passed a leak scan. Of \scXcN{} recordings labelled by both flight cuts, \scXcSame{} contain identical passes.
\begin{table}[h]\centering\footnotesize\caption{The corpus up to November 2026, with indented rows part of the row above. Where no receiver saw an aircraft within 50~km that day, passes are unknown, not zero.}\label{tab:corpus}
\begin{tabular}{lr}\toprule
Listed recordings & \scListed\\
Labelled recordings & \scRecordings\\
\quad with a pass & \scRecPass\\
\quad with no aircraft archived within 50~km that day & \scNoCov\\
Passes & \scPasses\\
\quad in an assumed 300~s window & \scAssumedWin\\
Sources & \scSources\\
Days labelled, of days with listings & \scDays{} of \scListedDays\\
Recordings with public audio & \scPublic\\
\quad with a pass & \scPubRecPass\\
\quad their passes & \scPubPasses\\
\bottomrule\end{tabular}\end{table}

\section{Field Tests in Detail}\label{sec:a-tests}
\input{build/tab_mainmore.tex}
\paragraph{Baselines.} DOPPLERGEIST~\citep{dopplergeist2026} runs at commit abb7d6d, dopplerprint~\citep{dopplerprint2026} at 0.1.0 and source 447dbed, and Unheard at commit 55936de, unmodified. Later versions may score differently. dopplerprint refuses input outside the contiguous US. On A2O it runs post hoc through a logged adapter that changes only the map projection, geoid and coverage area. It uses our audio, flight cuts, squares, start and slack, with GLO-30 terrain. Near midnight UTC it asks for the next day's flights. The blind input lacks them, so it gets an empty day. One batch 3 packet reached its hour limit. This deviation was decided from run statuses alone. DOPPLERGEIST was not run on A2O.
\paragraph{A2O batch 3.} The sites share none with earlier batches or the clock recorders. The scoring rule was fixed before the batch existed. Our detector, hashed before the batch was pulled, keeps clips with a clear pass. It tracks tonal ridges at 2~kHz in 1~s frames. A clear pass falls by at least 4\% over 80\% of its steps, has median contrast 8 or more and lasts at least 12~s. A hit is within 1~km of the position A2O serves. A2O's jitter rule moves that point up to \jbMaxM~m from the stored site. At one batch 1 site it is \jbGpsKm~km from the recorder's own GPS. The offset can move answers across the 1~km edge either way. A2O's code~\citep{bawserver} states a looser uncertainty of \jbTotOffM~m plus an assumed \jbTotGpsM~m GPS error, \jbTotM~m in all. Custom offsets remain unknown. If every hit within that total of the edge were lost, the rerun would keep \jbTotRefWorstK{} of the reference's \aoBThreeK{} hits and \jbTotDpWorstK{} of dopplerprint's \aoBThreeDpK, at the same places. An amendment from before the batch adds an early look and a second machine. Every row had landed before that look and none changed. Two looks at most double the $p$, to \aoBThreeTwoLookP. The reference answers \aoBThreeAns{} packets at a median \aoBThreeMedKm~km. Without the recorder with the most hits, \aoBThreeOutK{} hits at \aoBThreeOutSitesHit{} recorders remain. Post hoc, \aoBThreeHitEdge{} of the hits lie on the 60~s bound. In one draw of a packet per recorder, \aoBThreePsK{} of \aoBThreePsN{} packets hit against \aoBThreePsExp{} expected, and \aoBThreePsZeroPct\% of such draws give none. The rates thus rest on many recordings per recorder. A post hoc rerun on a second machine of the packets one machine answered keeps all \aoBThreeRerunHits{} of their hits, with none new. The \aoBThreeRerunUnans{} packets that machine left unanswered were not rerun.
\paragraph{Batch 4.} Batch 3's rule, squares, controls and scorer apply to \bFourSites{} A2O recorders not used before. They sit at \phFourPlN{} places, \phFourPlUsed{} used in earlier batches. A dated commit planned the batch before any audio was fetched. The verifier drew the squares and scored. The runner never held a position. Before any run the runner once read a note naming a few candidate sites and once saw a few site identifiers. The reference's hits fall at \bFourRefSitesHit{} recorders against \bFourSitesExp{} by chance ($p=$~\bFourRefP). Without its top recorder, $p=$~\bFourRefPLoo. dopplerprint and Unheard are within chance ($p=$~\bFourDpP{} and \bFourUnhP). Post hoc, the reference hits \phFourRefMoK{} of \phFourMoN{} moments ($p=$~\phFourRefMoP) and \phFourRefPlK{} of \phFourPlN{} places ($p=$~\phFourRefPlP). Under either control none place any. With so few places this checks only the direction of the result.
\paragraph{A2O batches 1 and 2.} These development batches use code tagged a2o-v1, registered and timestamped before scoring. The first batch's start is known to 5~s. Its hits give $p=$~\aoLocSitesP{} by site. Of them, \aoLocTopSiteK{} fall at one site. The second batch gives \aoLocBTwoAns{} answers and no hit. A site's squares together narrow it to a few hundred metres, so blindness rests on the frozen code reading one packet at a time. Those running the locator heard the sites' group names before the blind set was redrawn. Post hoc, with each start moved two hours later, the first batch hits at \aoLocCtlSites{} recorder, where the registered null expects 0.47 ($p=$~\aoLocCtlP). Post hoc, shifting every first-batch packet's start \aoShiftMin~min either way leaves \aoFullShiftK{} hit within 1~km. Shifts of 30~min give \aoFullLongK{} hit, and shifts of 1~min give \aoFullSixtyK{} against \aoFullSixtyExp{} by chance. These counts combine the two shift directions. Moving only the hits' start by \aoShiftMin{} or 30~min leaves \aoShiftK{} and \aoShiftLongK. With the start fixed at the label, \aoSlackZeroK{} of the \aoSlackZeroN{} hits remain. The 60~s window was chosen after a clock note informed by the sites. In it the first batch gives \aoSlackK{} hits against \aoSlackExp{} by chance. The registered run kept its registered slack.
\paragraph{Eclipse Soundscapes.} The registration is a git commit 42 min before the first run and before any Eclipse result. It was not timestamped. Its task list, not its text, sets the pooled test's threshold of two clear clips per recorder-day. Each clip runs in the site's 20~km square with the start known to a minute. A separate verifier scores against its own copy of the site places. The sites' places are public and were readable in our working files. The clock study's files held \eclClockInTask{} of the pooled test's sites, including \eclClockHit{} with a hit. The frozen code has no path to those files. Excluding those sites and scoring against each answer's share of its square within 1~km, pooled days still hit at \eclCleanSitesHit{} of \eclCleanSites{} sites ($p=$~\eclCleanSitesP), and single clips at \eclCleanClipSitesHit{} of \eclCleanClipSites{} ($p=$~\eclCleanClipSitesP). Post hoc on \eclCmpDgN{} clips, DOPPLERGEIST places \eclCmpDgK{} within 1~km and the reference \eclCmpDgOursK. dopplerprint's release places \eclCmpDpPypiK{} and its source \eclCmpDpSrcK.
\paragraph{Later xeno-canto.} The test is registered before any run, with code frozen at ecl-v1 and 8~kHz copies. Crashes and missing flight data count as misses. The registered prediction that hits increase with year fails ($p=$~\benchYrP). On an 8~kHz copy, \benchYrBridgeSame{} of \benchYrBridgeN{} earlier test cases keep the same answer, short of the registered \benchYrBridgePred.
\paragraph{Flight coverage.} ADS-B Exchange's \flAdsbxDays{} free days~\citep{adsbxsample} hold \flAdsbxRatio{} times adsb.lol's low, slow aircraft-minutes over the US.

\optfig[!htbp]{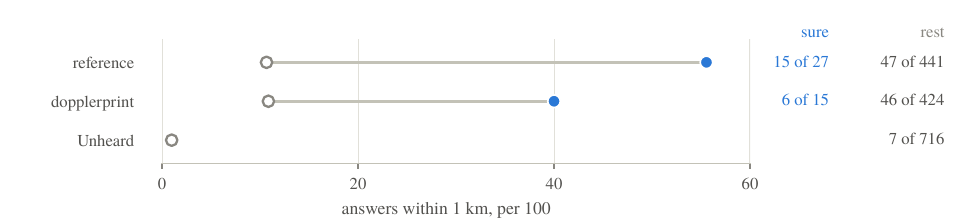}{Post hoc answers within 1~km per 100, split by each entry's confidence grade. The tests differ by entry, so the plot does not compare entries.}{fig:worth}
\paragraph{Confidence.}\label{sec:a-conf} \Cref{fig:worth} pools each entry's own locate tests. Solid marks show sure answers, open marks other answers and blanks no sure answer. The reference pools A2O batches 1 to 3, Eclipse clips, held-out and later xeno-canto, and US iNaturalist. dopplerprint pools its first test-set run, batch 1 and Eclipse. Unheard pools its three date-only runs on batches 1 and 3 and Eclipse. The reference marks a hit chance of 0.9 or more as sure. dopplerprint uses its high grade. Unheard sets its confident flag on none of these answers. The reference's confidence ranks its test-set answers (AUC \anAuc{} over \anAucN) but overstates them. Of the \confHiK{} sure hits before the test set, \confHiAoK{} are from A2O's first batch. On the later xeno-canto recordings it ranks them backwards (AUC \confXcyAuc).
\paragraph{What makes a hit.}\label{sec:a-hit} These checks are post hoc and exploratory, with no correction for multiple tests. Each test-set packet is described by the ADS-B aircraft closest to the true place. No entry hits on any of the \hmBThreeJetN{} packets whose closest aircraft is a jet, or the \hmBThreeFarN{} whose closest aircraft stays beyond 10~km. The reference hits \hmBThreePropK{} of \hmBThreePropN{} packets whose closest aircraft is a propeller aircraft and \hmBThreeRotK{} of \hmBThreeRotN{} a helicopter. Another \hmBThreeNoAc{} have no ADS-B sample. In the random sample, the reference places none of the \hmPrNoDetN{} recordings in which its detector finds no pass.
\paragraph{Square and slack.}\label{sec:a-sweep} Post hoc, the frozen reference and dopplerprint rerun the test set with one prior changed (\cref{tab:sweep}). The runs were written down before running and do not test the method. A start free within 5~s rather than 60~s increases the hits. At 300~s they fall. In 10~km squares both still beat a uniform guess on moments. In 5~km squares neither beats the full-disk chance. At 300~s the reference and dopplerprint hit \swRefSlLongPlK{} and \swDpSlLongPlK{} of the \phThreePlN{} places, and \swDpLongTo{} dopplerprint rows ran out of time and count as misses. \Cref{tab:sweep} gives hits within 1~km of \swN{} packets, packets answered, and moments hit of \phThreeMoN{} with $p$ against uniform guesses.
\begin{table}[h]\centering\footnotesize\caption{The test set rerun post hoc with one prior changed. For smaller squares chance is an upper bound.}\label{tab:sweep}
\begin{tabular}{@{}lrrrrrrrrr@{}}\toprule
& & \multicolumn{4}{c}{reference} & \multicolumn{4}{c}{dopplerprint}\\\cmidrule(lr){3-6}\cmidrule(l){7-10}
Square, slack & chance & hits & answered & moments & $p$ & hits & answered & moments & $p$\\\midrule
20~km, 60~s (as run) & \swOrigExp & \swRefOrigK & \swRefOrigAns & \swRefOrigMoK & \swRefOrigMoP & \swDpOrigK & \swDpOrigAns & \swDpOrigMoK & \swDpOrigMoP\\
20~km, 5~s & \swSlFiveExp & \swRefSlFiveK & \swRefSlFiveAns & \swRefSlFiveMoK & \swRefSlFiveMoP & \swDpSlFiveK & \swDpSlFiveAns & \swDpSlFiveMoK & \swDpSlFiveMoP\\
20~km, 300~s & \swSlLongExp & \swRefSlLongK & \swRefSlLongAns & \swRefSlLongMoK & \swRefSlLongMoP & \swDpSlLongK & \swDpSlLongAns & \swDpSlLongMoK & \swDpSlLongMoP\\
10~km, 60~s & \swSqTenExp & \swRefSqTenK & \swRefSqTenAns & \swRefSqTenMoK & \swRefSqTenMoP & \swDpSqTenK & \swDpSqTenAns & \swDpSqTenMoK & \swDpSqTenMoP\\
5~km, 60~s & \swSqFiveExp & \swRefSqFiveK & \swRefSqFiveAns & \swRefSqFiveMoK & \swRefSqFiveMoP & \swDpSqFiveK & \swDpSqFiveAns & \swDpSqFiveMoK & \swDpSqFiveMoP\\
\bottomrule\end{tabular}\end{table}

\section{Bird Prior}
\label{sec:a-birds}\label{sec:m-birds}\label{sec:res-birds}
BirdNET~\citep{kahl2021birdnet} detects a recording's species. Its geographic model, BirdNET 2.4~\citep{kahl2025birdnetv24}, read through the birdnet package~\citep{birdnetpy2026}, with geographic model 3.0~\citep{kahl2026geomodel} run as a comparison, gives each species' weekly occurrence on a 0.25\textdegree{} grid. For each detected species, its occurrence is mixed with the US mean in proportion to the detection's confidence. A cell's score sums the logs of these mixtures and adds a prior on where recordists record~\citep{phillips2009sample}. A temperature turns the score into a probability map. The region is the smallest set of cells with a quarter of its probability. From the date alone, a search of that region places \eclTTwoK{} of \eclTTwoSiteDays{} Eclipse site-days within 1~km. The region holds the recorder only on a day without two clear passes. This tests the prior, not the aircraft search.

\section{Other Archives and Europe}
\label{sec:a-global}
The locator and the frozen check run on research-grade iNaturalist~\citep{inaturalist} phone recordings from Europe and the US, and on xeno-canto from six European countries. The locator is the version frozen for Eclipse. iNaturalist coordinates are public, with stated uncertainty up to 250~m. Flight tracks are cut 30~km around the stated place. Unlike every other test's, this cut is centred on the truth. With the time known, the reference places \glIneuK{} of \glIneuN{} European phone recordings within 500~m against \glNullIneuExpH{} by chance. It places \glXceuKm{} of \glXceuN{} European xeno-canto ones within 1~km against \glNullXceuExpK.

At the stated place, the check confirms \glAuInusK{} of \glAuInusN{} US and \glAuIneuK{} of \glAuIneuN{} European phone recordings and \glAuXceuK{} of \glAuXceuN{} European xeno-canto ones. At decoys 150~km away, \glAuInusDec{} of the US phone recordings, \glAuIneuDec{} of the European phone recordings and \glAuXceuDec{} of the European xeno-canto recordings receive a decision. Of the last, \glAuXceuDecConf{} are confirmations. On European xeno-canto it confirms only slightly more often at stated places than at decoys. Shifts at stated places are no more frequent. The check's code was frozen \auditFreezeLagMin{} minutes after these and xeno-canto's decoys were first scored, with code and settings unchanged in between. The last \glAuXceuTwoN{} European xeno-canto recordings came after the freeze.

\paragraph{Park archive.} The test is blind and retrospective on \zenN{} recordings from \zenSites{} sites of a forest-soundscape archive~\citep{panwar2025forest}. The \zenUnbuilt{} without flight data count as misses. With the wrong time the locator places \zenCtlK{} of \zenCtlN{}. With a 0.2\textdegree{} square it places \zenLtK{} of \zenLtN. Public metadata name the park, and some sites, from the start time. The site table was world-readable for hours, so blindness depends on the frozen code.

\section{Licences}
\label{sec:a-licences}
Our code is under MIT.
Flight-track cuts from adsb.lol remain under ODbL~1.0~\citep{adsblol}. OpenSky data, for non-profit research and education only~\citep{openskyterms}, are in no reported number. ADS-B Exchange's free days are for testing and evaluation~\citep{adsbxsample}. adsb.fi's~\citep{adsbfi} are for personal, non-commercial use only. None is redistributed. ADS-B Exchange is used only for the coverage ratio in \cref{sec:a-tests}. xeno-canto audio is not copied: each recording is referenced by catalogue number under its own Creative Commons licence and credited to its recordist~\citep{xenocanto}. The Livermore recordings~\citep{pathak2026livermore} and LibriSpeech~\citep{panayotov2015librispeech} are under CC~BY~4.0 licences. GLO-30 is used under the Copernicus DEM licence~\citep{glo30}. BirdNET 2.4's record states CC~BY-NC~4.0~\citep{kahl2025birdnetv24}. The geographic model 3.0's record states Apache~2.0~\citep{kahl2026geomodel}. The birdnet package states CC~BY-NC-SA~4.0 for the models it provides~\citep{birdnetpy2026}. All are used for non-commercial research only. The park archive~\citep{panwar2025forest} and the Eclipse Soundscapes records~\citep{eclipsesoundscapes,eclipse2026esid278} are CC~BY~4.0. The random sample's audio is read from its public archives, including Ecosounds~\citep{ecosounds}, for scoring and analysis only. None of it is redistributed. Eclipse Soundscapes and Acoustic Observatory recorders sit outside on volunteers' or public land and can capture passers-by. Their audio is read only for aircraft passes and birds. None of it is redistributed.

\section{Suppressor and Cleaner Settings}
\label{sec:a-defense}
On held-out xeno-canto windows without speech, FFmpeg's filters, the phone path and DeepFilterNet3 retain most passes the detector finds unprocessed. RNNoise and Unheard's tone remover~\citep{unheard2026} retain few or none (\cref{tab:defense}). A pass counts as found only where the detector finds a line at its frequency. The suppressors are RNNoise~\citep{valin2018rnnoise}, FFmpeg's arnndn with the community models cb and sh~\citep{richards2018rnnoisemodels}, DeepFilterNet3~\citep{schroeter2023deepfilternet3} release 0.5.6~\citep{dfnrelease} and FFmpeg's afftdn at default and strong settings~\citep{ffmpegfilters}. The phone path is 8~kHz Opus~\citep{rfc6716} at 16~kb/s. Inputs are resampled to 48~kHz for the neural suppressors and back. Unheard's tone remover runs as published.

Our cleaner\label{sec:m-cleaner} joins spectral peaks into gliding tracks and refills those above the background with background-level noise. It runs offline or live with lookaheads of 0, 1, 3 and 5~s. Under a rule registered before the run, the offline cleaner is applied to \defLocN{} earlier time-known reference hits. The uncleaned reruns retain every hit. After the cleaner \defLocK{} survive, \defLocEclK{} of \defLocEclN{} and \defLocAoK{} of \defLocAoN{} by test. The fraction is not a hit rate. Its settings file was written after its first output. No informed attacker was tested.
\input{build/tab_defense.tex}
Of the \defDpLivN{} Livermore clips dopplerprint places within 1~km, \defDpLivNoneK{} still hit at 8~kHz. RNNoise retains \defDpLivRnK{} of those hits, DeepFilterNet3 \defDpLivDfnK{}, added speech alone \defDpLivMixK{} and our offline cleaner \defDpLivOffK. These clips are not fully held out, since the cleaner was spot-checked on Livermore clips.

%% file: build/tab_audit.tex
\begin{table}[htbp]\centering\small
\caption{Verdicts of the frozen clock check, per recording and clock shift. Right: the verdict picks the true shift, scorable only on Livermore.}\label{tab:audit}
\begin{tabular}{lrrrr}\toprule
Set & Cases & Confirmed & Shifted & Right\\\midrule
Livermore, true clock & 39 & 18 & 0 & 18\\
Livermore, clock shifted & 195 & 0 & 19 & 19\\
xeno-canto, detector finds a pass & 715 & 18 & 1 & \\
\quad expected by chance, from its decoys & 715 & 1.0 & 5.0 & \\
\midrule
Held-out decoys, pooled, observed & 908 & 1 & 7 & \\
\bottomrule\end{tabular}
\end{table}

%% file: build/tab_controls.tex
\begin{table}[t]\centering\footnotesize
\caption{The reference's controls on the test set's first run: hits, recorders with a hit, chance and $p$, counting recorders as independent as registered.}\label{tab:controls}
\setlength{\tabcolsep}{3.5pt}\begin{tabular}{@{}llrrrr@{}}\toprule
Arm & Plan & Answers & Hits (recorders) & Chance (recorders) & $p$\\\midrule
Stated time & registered & 98 & 21 (13) & 2.7 & $2\times10^{-7}$\\
Start stated 5~min late & registered & 46 & 1 (1) & 2.7 & 0.94\\
Start stated 5~min early & registered & 61 & 2 (2) & 2.7 & 0.76\\
Start stated 2~h late & registered & 31 & 0 (0) & 2.7 & 1\\
Another site's audio & post hoc & 73 & 1 (1) & 2.7 & 0.94\\
Square's centre & post hoc & 360 & 1 (1) & 2.7 & 0.94\\
\bottomrule\end{tabular}
\end{table}

%% file: build/tab_protocol.tex
\begin{table}[h]\centering\footnotesize
\caption{When each plan was registered and timestamped, relative to its test's first run, and how often it was amended. * the stamp covers the scoring rule or sealed squares.}\label{tab:protocol}
\begin{tabular}{llllrr}\toprule
Test & Registered & Timestamped & Pre-registered & Amendments & After first run\\\midrule
Eclipse, time known & 42 min before & git only & no & 5 & 4\\
Eclipse, date only & 2.2 h before & git only & no & 10 & 5\\
A2O batch 1 & 23 min before & 2.3 h after & no & 8 & 6\\
A2O batch 2 & 1.8 h before & 2 min after & no & 2 & 2\\
A2O batch 3 & 75 min before & 72 min before* & yes & 13 & 7\\
Later xeno-canto & 45 min before & 44 min after & no & 3 & 2\\
Live streams & 60 min before & 4.4 h after & no & 8 & 4\\
Park archive & 2.0 h before & 7 min after & no & 2 & 1\\
Park archive, 0.2\textdegree{} square & 6.4 h before & 4.3 h before & yes & 2 & 0\\
Unheard, batch 1 and Eclipse & 33 min before & 33 min before & yes & 3 & 3\\
Unheard, batch 3 & 2.6 h before & 2.6 h before* & yes & 2 & 0\\
Cleaner vs the reference & 15 min before & 15 min before & yes & 0 & 0\\
Random public recordings & 72 min before & 64 min before & yes & 3 & 1\\
New Eclipse places & 2.8 h before & 2.8 h before & yes & 2 & 0\\
\bottomrule\end{tabular}
\end{table}

%% file: build/tab_mainmore.tex
\begin{table}[h]\centering\small
\caption{Further rows of \cref{tab:main}, counted the same way. $\dagger$ post hoc.}\label{tab:mainmore}
\setlength{\tabcolsep}{5pt}\renewcommand{\arraystretch}{0.92}\begin{tabular}{@{}llrrrrr@{}}\toprule
Set & Entry & Recordings & Answered & Hits & Chance & Recorders \\
\midrule
Development, batch 1 & Unheard, date only & 295 & 268 & 5 & 0.09 & 3\\
 & square's centre\rlap{\smash{$^\dagger$}} &  & 295 & 2 & 2.32 & 2\\
\addlinespace[1.5pt]
Development, Eclipse & Unheard, date only & 187 & 139 & 0 & 0.06 & 0\\
 & square's centre\rlap{\smash{$^\dagger$}} &  & 187 & 0 & 1.47 & 0\\
\addlinespace[1.5pt]
Live streams & reference & 80 & 63 & 0 & $<$0.01 & 0\\
\bottomrule\end{tabular}
\end{table}

%% file: build/tab_defense.tex
\begin{table}[h]\centering\footnotesize
\caption{Passes the detector still finds in 35 held-out xeno-canto windows after each processing.}\label{tab:defense}
\begin{tabular}{lc}\toprule
Processing & Passes found\\\midrule
None & 30/35\\
ffmpeg afftdn & 28/35\\
ffmpeg afftdn, strong & 20/35\\
ffmpeg arnndn (cb) & 26/35\\
ffmpeg arnndn (sh) & 26/35\\
Opus 16 kb/s & 23/35\\
RNNoise & 2/35\\
DeepFilterNet3 & 18/35\\
Unheard & 0/35\\
Ours, offline & 0/35\\
Ours, live & 0/35\\
\bottomrule\end{tabular}
\end{table}